\documentclass[10pt, twocolumn]{IEEEtran}
\normalsize
\usepackage{amsthm,amssymb,amsmath,tikz,graphics,cite,float,graphicx,epstopdf,epsfig,verbatim,url}
\usetikzlibrary{automata}
\usetikzlibrary{shapes,arrows}

\usepackage[utf8]{inputenc}

\usepackage{capt-of}
\usepackage[noend]{algpseudocode}
\usepackage{algorithmicx}
\usepackage[ruled]{algorithm}

\usepackage{lipsum}

\usepackage{enumerate}
\usepackage{mdwtab}
\usepackage{subfigure}
\usepackage{placeins}
\usepackage{setspace}

\allowdisplaybreaks

\theoremstyle{plain}
\theoremstyle{remark}

\begin{document}
\newtheorem{theorem}{Theorem}
\newtheorem{lemma}{Lemma}
\newtheorem{conjecture}{Conjecture}
\newtheorem{corollary}{Corollary}
\newtheorem{definition}{Definition}
\newtheorem{property}{Property}
\newtheorem{remark}{Remark}
\newtheorem{proposition}{Proposition}

\bibliographystyle{plain}

\title{Fundamentals of Power Allocation Strategies for Downlink Multi-user NOMA with Target Rates}


\author{Jos\'{e}~Armando~Oviedo,~\IEEEmembership{Member,~IEEE,}
        and~Hamid~R.~Sadjadpour,~\IEEEmembership{Senior Member,~IEEE}
\thanks{Manuscript received January 29, 2019; revised July 6, 2019 and October 6, 2019; accepted December 4, 2019. The associate editor coordinating the review of this article and approving it for publication was D. R. Brown. (Corresponding author: José Armando Oviedo.)}
\thanks{Copyright (c) 2019 IEEE. Personal use of this material is permitted. However, permission to use this material for any other purposes must be obtained from the IEEE by sending a request to pubs-permissions@ieee.org.}%
\thanks{The authors are with the Department of Electrical and Computer Engineering, University
of California, Santa Cruz, CA 95064 USA. (e-mail xmando@soe.ucsc.edu, hamid@soe.ucsc.edu.)}}

\maketitle

\IEEEpeerreviewmaketitle

\begin{abstract}
For downlink multi-user non-orthogonal multiple access (NOMA) systems with successive interference cancellation (SIC) receivers, and a base-station not possessing the instantaneous channel gains, the fundamental relationship between the target rates and power allocation is investigated. It is proven that the total interference from signals not removed by SIC has a fundamental upper limit which is a function of the target rates, and the outage probability is one when exceeding this limit. The concept of well-behaved power allocation strategies is defined, and its properties are proven to be derived solely based on the target rates. The existence of power allocation strategies that enable NOMA to outperform OMA in per-user outage probability is proven, and are always well-behaved for the case when the outage probability performance of NOMA and OMA are equal for all users. The proposed SIC decoding order is then shown to the most energy efficient. The derivation of well-behaved power allocation strategies that have improved outage probability performance over OMA for each user is outlined. Simulations validate the theoretical results, demonstrating that NOMA systems can always outperform OMA systems in outage probability performance, without relying on the exact channel gains.
\end{abstract}


\section{Introduction}
Due to the rapidly increasing demand for higher data-rates, more connected users and devices, and diversity of deployments, power-domain non-orthogonal multiple access (NOMA) is being sought to help improve the capacity and user multiplexing of downlink (DL) and uplink cellular systems \cite{NOMA5GBeyond}. The 3rd Generation Partnership Project has already conducted the study items for both downlink NOMA for LTE-Advance \cite{3GPP:MUST}, and for uplink NOMA in New Radio (NR) \cite{3GPPNOMA}. With the attention that NOMA receives from the academic, private, and standard sectors, it is only a matter of time before NOMA is implemented in future wireless system deployments. 

Power-domain NOMA takes advantage of superposition coding (SC) and received signals with disparate received powers, and employs a successive interference cancellation (SIC) enabled receiver in order to successively remove interference according to the ordered received powers of the signals. In the downlink, the base-station (BS) transmits the signals to multiple users over a common transmission period and bandwidth. Each user then obtains its own signal by employing SIC to decode and remove the interference of signals with greater power than its own. NOMA is in contrast to orthogonal multiple access (OMA), which assigns non-overlapping time slots (or frequency sub-bands) to each user in order to avoid interference. Since channel capacity increases linearly with time and frequency, and only logarithmically with transmit power, NOMA can outperform OMA in terms of achieving higher data rates \cite{InfTh:CT}. 

Although it is proven in \cite{FairNOMAFull} that there always exists a power allocation approach for NOMA that can outperform OMA for the general multi-user NOMA case in terms of information capacity, this power allocation strategy relies on the BS having perfect instantenous channel state information (CSI) at the transmitter, i.e. the exact channel gain value, which is not a realistic assumption in wireless system deployments. This is due to the limitations of the CSI formats that are fed back by the users to the BS, and time gaps between channel estimation by the BS and the associated downlink transmissions. In 4G and 5G system deployments \cite{Dahlman5G}, the BS determines whether a target rate can be supported based on realistic CSI formats (rank indicator, precoding matrix indicator, channel quality indicator, etc.), and selects the remaining transmission parameters which will accommodate the downlink transmission at the indicated target rate.
Therefore, it is important that a DL NOMA system be able to determine the power allocation for all NOMA users based on the available information in real system deployments, and not the exact channel gain values.

In this work, it is assumed that the BS has determined that the target rates can be supported by the channels for all users, but that it does not possess the full CSI in the form of the channel gains, and must determine the NOMA power allocation strategy based on the available information. For the baseline OMA system which NOMA will be compared to, a general TDMA approach is used with each user being allocated a fraction of the total transmit time duration, and the BS uses full transmit power for each transmission. For the NOMA system, the transmit power allocation strategy and associated SIC decoding order are completely determined by the users' target data rates and the associated OMA transmit time durations. It is proven that a power allocation strategy must be such that the received interference coefficient for each signal must be below a fundamental threshold in order to not experience unavoidable outages. The concept of {\it well-behaved} power allocation strategies is defined, and shown that these strategies satisfy the interference requirement. 
It is then proven that there always exists a power allocation strategy such that all users will have NOMA outage probability performance equal to that of OMA, that the proposed SIC decoding order is the most energy efficient, and that such a strategy is always well-behaved. Finally, the approach to derive a well-behaved power allocation strategy such that a user can achieve better outage probability performance with NOMA over OMA is outlined.
 
\section{Previous Work and Current Contribution} \label{sec:previous}

The outage probability of NOMA was investigated in \cite{NOMA-HARQ:Choi}, where multiple users transmit simultaneously to multiple receivers using a uniform power allocation approach, and it was shown the outage probability is improved when NOMA is combined with H-ARQ vs OMA with H-ARQ. 
The authors in \cite{NOMA-RDU:DYFP} showed that the power allocation and interference coefficients of each user are fundamentally dependent on the particular user's required rate, and thus the wrong selection of coefficients can lead to an outage with probability equal to 1. 
The usage of NOMA in a cognitive-radio inspired approach was mentioned in \cite{5GNOMA:DFP}, where a user with weak channel condition is seen as the primary user and is provided as much power as needed in order to achieve its minimum rate, and the user with stronger channel is treated as the secondary user and receives any remaining power not allocated to the weaker user, and the outage probability of both is shown to clearly depend on pairing users with stronger channels. 

A couple of works have focused on utilizing the rate achieved using OMA as the minimum rate required for NOMA, and the associated power allocation solution which achieves this condition.
The region of power allocation coefficients that allow NOMA to outperform OMA in the downlink is first defined for the two-user case in \cite{FairNOMA:Infocom}. 
The authors in \cite{DNOMA:YDFA} then use a power allocation approach in this region to analyze the outage performance and diversity orders of two paired users, according to their relative channel gains, and extend the work to the uplink case. 
In \cite{FairNOMAFull}, the power allocation coefficients for a multi-user NOMA system which always outperforms OMA are proven to always have a sum less than or equal to 1, and hence a valid power allocation strategy for NOMA always exists that outperforms OMA in terms of capacity, while using less power than OMA. 
The work in \cite{FairRateComp:ZengDobre2017} extends the concept of power allocation fairness with regards to NOMA compared to OMA, showing there always exists a power allocation for NOMA that allows the rate to outperform the rates of the generalized FDMA case with optimizing resource allocation.
In \cite{CapacityComp:ZengDobre2018}, the authors prove that  for any power  and resource allocation in FDMA, there always exists a power allocation strategy that will provide a superior sum-rate and ergodic rate for NOMA over OMA, while developing a user admission scheme to maintain a balance between the number of total admitted users and sum-rate performance.

More recent works have focused on optimizing the power allocation strategy. 
The authors in \cite{JPNOMA:ICC2016} propose a joint optimization of user pairing and power allocation by optimizing a cost function dependent on the instantaneous achievable rates and a metric based on proportional fairness. The scheduling and power allocation algorithm that solves the optimization problem is compared to the fractional transmit power control algorithm and shown to improve performance for the user with stronger channel, while performance is not always improved for the user with weaker channel. 
The work in \cite{CognitiveNOMA:ZengDobre2016} uses a new algorithm to solve the cognitive radio NOMA power allocation problem which can outperform the fractional transmit power algorithm for admitting secondary users into the network. 
In \cite{On-OptPA-NOMA-QoS}, the authors seek to optimize the sum-rate of a multi-user downlink NOMA system by using a constraint based on the total power allocated to the signals at each SIC stage, and its relation to the minimum required rate for each signal to be decoded. 
The authors in \cite{OptPA-NOMA} studied several algorithms that solve the NOMA power allocation problem, and point out that not many existing works had considered the strict constraint for the power allocations to follow the order of SIC decoding in their algorithms. They proposed to incorporate the matching algorithm and optimum power allocation, and found that the constraint has a significant impact on the power allocation solution, which also yields superior performance over existing schemes. 
In \cite{XiaoZhuChoi:PA-BF-NOMA5Gmm2018}, a new solution is proposed for a system that clusters users in order to solve the joint beamforming and power allocation problem by breaking the problem up into two separate sub-problems, where the goal is to maximize the sum-rate of each cluster. 
In \cite{ResourcePA-OPt}, the authors propose a joint resource (bandwidth) and power allocation approach that optimizes a cost function that is an affine function of the power allocations and bandwidths. It is demonstrated that this algorithm can outperform the approach of simply optimizing the power allocations with fixed bandwidths, as well as the baseline OMA approach.
The work in  \cite{YanYanDingNg:BF-PA-NOMA-MIMO} proposes a joint beamforming and power allocation solution to a coordinated multi-point MIMO-NOMA system, where the intra-cell interference between clusters is cancelled through transmit beamforming, and the power allocation is designed to maximize one user's rate, while maintaining the rate of the second user.
In \cite{DynamicPA-MA}, the authors derive a weighted sum-rate maximization algorithm to find the power allocation per subcarrier for a pair of users, and show that the performance of their approach improves as the diversity order of the system increases. 
In \cite{PA-CoMP-NOMA}, the authors study the power allocation approach for multi-tiered cellular networks with cell-center users and a cell-edge user who is eligible for coordinated multi-point transmission. A joint power optimization algorithm is formulated, including target rates for each user, and due to the prohibitive complexity, a distributed power optimization problem is formulated and shown to exhibit near optimum performance. The constraint on the power allocation coefficients relies on a linear function derived from the SINR for each SIC stage. 

In the previous works, the power allocation constraints either do not consider the necessary requirements for successful SIC performance, or use constraints that do not give the fundamental relationship between power allocation and outage, such that an outage event is certainly avoided.
In other words, the target rates and power allocation required to ensure whether successful SIC performance is even feasible at each stage of SIC decoding for each user is not {\em directly} considered, and this can cause unnecessary unavoidable outages to occur for multiple users. In fact, this phenomenon was described and demonstrated in \cite{CANOMA} for two-user cache-aided NOMA systems, and in \cite{PA-Limits-DLNOMA} for multi-user downlink NOMA systems with a QoS constraint. 

The main contribution of this work is to provide a comprehensive theoretical treatment of power allocation strategies, and how they are related to the SIC decoding order selected, the associated target rates of the users, and any associated OMA parameters that affect the design of the NOMA power allocation coefficients, while not relying on the channel gain value. In particular, this work provides the following:
\begin{itemize}
	\item The fundamental maximum interference that a particular signal can tolerate from other NOMA users before its outage probability is equal to 1, regardless of how strong the channel SNR gains are;
	\item The definition of a {\em well-behaved} power allocation strategy, which causes the outage thresholds to be lesser for the signals of users that are earlier in the SIC decoding order. This condition is then shown to always lead to an acceptable value of NOMA interference;
	\item The baseline power allocation strategy which achieves outage performance equal to that of OMA for all users is derived in closed-form, and is used to prove that the proposed decoding order is energy efficiency optimal; 
	\item The baseline power allocation strategy is used to derive the conditions for acquiring a power allocation strategy where all users have superior NOMA outage performance over OMA, the exact approach for increasing the power allocation beyond the baseline strategy is outlined in detail, and a quick example of a power allocation strategy that satisfies all of these conditions is presented along with its performance. 
\end{itemize}
The necessity of such results in further studies of NOMA is clear, in the sense that when performing numerical studies of different algorithms applied to solve the power allocation problem for complex cellular deployments, 
the search space for the multi-user power allocation strategies can be greatly reduced to the subset of strategies that are well-behaved and improve the outage performance of NOMA over OMA.

\section{System Model}\label{sec:system}
Consider a wireless downlink system serving $K$ users. The BS will transmit $K$ multiplexed signals to the $K$ users. Let the signal for user $n$ be $x_n$, $1 \leq n \leq K$, such that $x_n$ is complex normally distributed with $\mathbb{E}[|x_n|]=1$, and is transmitted with transmit SNR $\xi$ through a wireless slow fading channel with SNR gain $|G_n|^2$. The channel gain $G_n$ can be one of many fading channels, such as a Rayleigh fading channel $|G_n|^2 \sim\mathrm{Exponential}(\beta_n)$, where the value $\beta_n$ can depend on the distance from the BS, or a MIMO fading channel $\boldsymbol{H}_n$ with precoding vector $\boldsymbol{p}$ at the transmitter and detection vectors $\boldsymbol{v}_n$ at the receivers, such that the overall channel SNR gain is $|G_n|^2 = |\boldsymbol{v}_n\boldsymbol{H}_n\boldsymbol{p}|^2$. The channel gain is not assumed to be known at the BS.

In the case of OMA, the general TDMA model is used. For a normalized total transmit time duration, the fractional time duration allocated to user $n$ is $\tau_n$, such that $\sum_{n=1}^K \tau_n = 1$. The received signal at user $n$ is given by $y_n = \sqrt{\xi}G_n x_n + z_n$, where $z_n\sim\mathcal{CN}(0,1)$ is the receiver thermal noise. Since user $n$ is allocated $\tau_n$ fraction of the total time resource, the capacity of user $n$ using OMA is given by $C_{n}^\mathrm{oma} = \tau_n\log_2\left(1 + \xi |G_n|^2 \right)$.

For the NOMA system, user $n$ has power allocation coefficient $a_n$, such that $\sum_{n=1}^K a_n = 1$. The received signal at user $n$ is
\begin{equation}
	r_n = G_n\sum_{l=1}^K x_l\sqrt{a_l\xi} + z_n.
\end{equation}
Using SIC, the receiver at user $n$, $n>1$, will decode the messages of users $m<n$ in ascending order, starting with $m=1$ (the SIC decoding order details are discussed in section \ref{sec:QoS}). Therefore, user $n$ will perform SIC on the signals of user $m=1,\ldots,n-1$, which have the form
\begin{equation}
    y_{n\rightarrow m} = G_n(\underbrace{x_m\sqrt{a_m\xi}}_\text{user-$m$ signal for SIC}+\sum_{l=m+1}^K x_l\sqrt{a_l\xi}) + z_n, 
\end{equation}
until it can obtain the intended signal for user $n$ given by 
\begin{equation}
	y_n = G_n(\underbrace{x_n\sqrt{a_n\xi}}_\text{user-$n$ signal}+\sum_{l=n+1}^K x_l\sqrt{a_l\xi}) + z_n,
\end{equation}
where $\sum_{l=n+1}^K x_l\sqrt{a_l\xi}$ are the signals that need not be decoded using SIC by user $n$ in order to decode its own signal, 
and thus are treated as interference.

For the power allocation coefficients $a_1,\ldots,a_K$, the capacity of the channel for user $n<K$ is
\begin{equation}
	C_n(a_1,\ldots,a_K)=\log_2\left(1 + \tfrac{a_n\xi |G_n|^2}{1+\xi |G_n|^2\sum_{l=n+1}^K a_l} \right),
\end{equation}
and user $K$ has capacity 
\begin{equation}
	C_K(a_1,\ldots,a_K)=\log_2\left(1 + a_K\xi |G_K|^2 \right).
\end{equation}

Meanwhile, for each user $n$ to achieve its capacity, it must have the capacity to decode the messages sent to all users $m<n$, and then subtract their signals from the composite signal received. The capacity of the channel which user $n$ will use to decode user $m$'s message is given by 
\begin{equation}
	C_{n\rightarrow m}(a_1,\ldots,a_K) = \log_2\left(1 + \tfrac{a_m\xi |G_n|^2}{1+\xi |G_n|^2\sum_{l=m+1}^K a_l} \right).
\end{equation}

\section{Basics of NOMA power allocation for systems with target rates} \label{sec:QoS}
Let each user $n$ have its information transmitted at a target rate $R_n$. First, define the event when user $n$ experiences an outage in an OMA system as
\begin{align}
	\mathcal{B}_n^\text{oma} = \left\{ C_n^\text{oma} < R_n \right\} = \label{eq:OMA_out}\left\{|G_n|^2 <\tfrac{2^{R_n/\tau_n}-1}{\xi}\right\}.
\end{align}
Since the goal of NOMA is to outperform OMA with respect to certain metrics (outage probability in this study), the OMA parameter $\tau_n$ and associated outage events affect the selection of the NOMA power allocation strategy. Considering that the OMA outage event can be normalized by dividing by $\tau_n$, yielding the normalized rate $\frac{R_n}{\tau_n}$, this normalized rate can be used as the quantity for determining the SIC decoding order. This quantity can also be seen from equation (\ref{eq:OMA_out}) as the determining factor for the value of the OMA outage threshold. So selecting the decoding order based on increasing OMA outage thresholds seems intuitive, since the NOMA outage thresholds will be directly compared to the OMA outage thresholds when finding the power allocation strategy.

Let the ordering of the user indices follow the ordering of the relationship $\frac{R_n}{\tau_n}$, such that indices $(1,\ldots,K)$ correspond to $\frac{R_1}{\tau_1}<\cdots<\frac{R_K}{\tau_K}$. A user $n=1,\ldots,K$, will experience an outage during the decoding process of its information if any of the following occurs:
\begin{equation}
	C_n(a_1,\ldots,a_K)<R_n\hspace{2mm}\text{ OR }\hspace{2mm}C_{n\rightarrow m}(a_1,\ldots,a_K)<R_m,
\end{equation}
for any $m<n$. Define the following events based on the specific signal which user $n$ needs to detect and decode, where $n=2,\ldots,K$, and $m<n$, 
\begin{align}
    \label{eq:outageB}&\mathcal{B}_n = \{C_n(a_1,\ldots,a_K)<R_n\}\\
     &\mathcal{B}_{n\rightarrow m}=\{C_{n\rightarrow m}(a_1,\ldots,a_K)<R_m\} \nonumber.
\end{align}
The NOMA outage event $\mathcal{B}_n^\text{out}$ at user $n$ can then be described as
\begin{align}
	&\mathcal{B}_n^\text{out} = \mathcal{B}_n\cup\left( \bigcup_{m=1}^{n-1}\mathcal{B}_{n\rightarrow m}\right).
\end{align}
Note that $\mathcal{B}_1^\text{out}=\mathcal{B}_1$ because user $1$ does not perform SIC in order to decode its own signal.

\subsection{Certain outage in NOMA transmissions}
From the definition of the NOMA outages, the following theorem is obtained.
\begin{theorem}\label{thm:power_limit}
    For a $K$-user DL NOMA system with user target rates $R_1,\ldots,R_K$ and power allocation coefficients $a_1,\ldots,a_K$, define $A_n=\sum_{l=n+1}^K a_l, \forall n=1,\ldots,K-1$, which is the interference coefficient in the received signal that users $n,\ldots,K$ will use to detect and decode user $n$'s information. If $\exists n$ such that $A_n > 2^{-\sum_{l=1}^n R_l}$, then for user $n$ and $\forall l>n$,
    \begin{equation}
        \mathrm{Pr}\{\mathcal{B}_n\} = \mathrm{Pr}\{\mathcal{B}_{l\rightarrow n}\} = 1,
    \end{equation} 
    and thus SIC will fail for all users $l=n,\ldots,K$. 
\end{theorem}
\begin{IEEEproof}
	See appendix \ref{proof:power_limit}.
\end{IEEEproof}
Theorem \ref{thm:power_limit} demonstrates that there is a fundamental relationship between the set of target rates $R_n$ and associated power allocation coefficients $a_n$, $n=1,\ldots,K$. It also demonstrates that as these target rates increase, the values of $a_n$ decrease rapidly, indicating that as the target rates increase for users earlier in the SIC decoding order, the amount of available power to the users later in the decoding order decreases.

Note that this does not indicate that the rate for user $n$ is guaranteed if $A_n<2^{-\sum_{l=1}^n R_l}$, since the total power allocation available to users $n,\ldots,K$ may be less than $2^{-\sum_{l=1}^{n}R_l}$ to begin with. In fact, a bound that is more case specific to the actually selected power allocation coefficients, as outlined in \cite{NOMA-RDU:DYFP}, is 
\begin{equation}\label{eq:interference_limit}
	A_{n-1}>A_n2^{R_n}, n=2,\ldots,K-1.
\end{equation} 
However, although it is a more strict bound, it is dependent on the specific case of power allocation coefficients, whereas the bound provided in theorem \ref{thm:power_limit} is a fundamental upper limit on the received interference coefficient that cannot be exceeded by any power allocation scheme. So a set of power allocation coefficients that satisfy equation (\ref{eq:interference_limit}) also satisfy theorem \ref{thm:power_limit}.

With the assumption that the power allocation coefficients are selected such that $A_n = \sum_{l=n+1}^K a_l < 2^{-\sum_{l=1}^{n}R_l}$, it should also be noted that in order for the SIC process to reach a decoding stage $n$, it does so with a certain probability at each user $k\geq n$. In other words, if users $k=n,\ldots,K$ are going to avoid an outage, they must sequentially decode messages $m=1,\ldots,n$ successfully in the process. Given the sequential nature of the decoding process, it is therefore desirable that the initial decoding stages have lower outage probabilities.

\subsection{Well-behaved power allocation strategies}
In light of theorem \ref{thm:power_limit}, the outage events for user $n=1,\ldots,K-1$ can be rewritten as 
\begin{align}
	&\mathcal{B}_n = \left\{|G_n|^2 < \tfrac{2^{R_n}-1}{\xi(a_n-(2^{R_n}-1)\sum_{l=n+1}^K a_l)} \right\}\\
	&\mathcal{B}_{n\rightarrow m} =\left\{|G_n|^2 < \tfrac{2^{R_m}-1}{\xi(a_m-(2^{R_m}-1)\sum_{l=m+1}^K a_l)} \right\},\nonumber
\end{align}
for $m=1,\ldots,n-1$, and $\mathcal{B}_K = \left\{|G_K|^2<\frac{2^{R_K}-1}{a_K\xi}\right\}$. This means that the overall outage event $\mathcal{B}_n^\text{out}$ can be expressed as 
\begin{align}
	\label{eq:decoding_event_channel}\mathcal{B}_n^\text{out} = &\left\{|G_n|^2 < \tfrac{2^{R_n}-1}{\xi(a_n-(2^{R_n}-1)\sum_{l=n+1}^K a_l)}\right\}\\
	&\cup\left(\bigcup_{m=1}^{n-1} \left\{|G_n|^2 < \tfrac{2^{R_m}-1}{\xi(a_m-(2^{R_m}-1)\sum_{l=m+1}^K a_l)} \right\} \right)\nonumber.
\end{align}
It is not desirable that a user $n$'s outage probability be primarily dictated by the success or failure of the earlier decoding stages. Since each event in equation (\ref{eq:decoding_event_channel}) is determined by a finite length interval in the form of $(0,\alpha)\subset\mathbb{R}^+$, it is clear that $\exists k, 1\leq k\leq n$, such that $\forall m=1,\ldots,n$,
\begin{align}
	&\tfrac{2^{R_k}-1}{\xi(a_k-(2^{R_k}-1)\sum_{l=k+1}^K a_l)}\geq \tfrac{2^{R_m}-1}{\xi(a_m-(2^{R_m}-1)\sum_{l=m+1}^K a_l)}, 
\end{align}
$\Rightarrow\mathcal{B}_n^\text{out} = \mathcal{B}_k$. This leads to the following proposition.
\begin{proposition}\label{prop:power_coeff_relate}
	For a $K$ user DL NOMA system with target rates $R_1,\ldots,R_K$, if the associated power allocation coefficients $a_1,\ldots,a_K$ are selected such that 
	$\mathcal{B}_n^\text{out} = \mathcal{B}_n, \forall n=1,\ldots,K,$
	then 
	\begin{align}\label{eq:noma_power_relate}
		a_{n}\geq a_{n+1} \tfrac{2^{R_{n+1}}(2^{R_{n}}-1)}{2^{R_{n+1}}-1}, n = 1,\ldots, K-1, 
	\end{align}
	and $A_n = \sum_{l=n+1}^K a_l< 2^{-\sum_{l=1}^n R_l}$, satisfying the requirement from theorem \ref{thm:power_limit}.
\end{proposition}
\begin{IEEEproof}
	See appendix \ref{proof:power_coeff_relate}.
\end{IEEEproof}
The above proposition provides the relationship between the power allocation strategies and the desired outage events. In other words, the outage probability to decode user $n$'s information should not be determined by an outage event during a SIC stage, but by the outage event of its own signal. Also, note that this condition also favors the decoding probability of all users whose signals are earlier in the SIC decoding order, as it places a lesser upper bound on the amount of NOMA interference received. From here on, any power allocation strategy which satisfies proposition \ref{prop:power_coeff_relate}, and by extension theorem \ref{thm:power_limit}, will be defined as being a {\em well-behaved} strategy.

The concept of {\em well-behaved} is not simply a preference, but an essential component for selection of an efficient NOMA power allocation strategy which aims to improve the outage performance over OMA for any user $n$ without having their performance sabotaged by an earlier SIC decoding stage $m$. For example, suppose $\exists m$ and $n$, $m<n$, for a non-well-behaved power allocation strategy such that $\mathcal{B}_n\subset\mathcal{B}_{m\rightarrow n}$, then $\mathcal{B}_n^\text{out} = \mathcal{B}_{m\rightarrow n}$. Furthermore, let the power allocation strategy be such that $\mathcal{B}_m = \mathcal{B}_m^\text{oma}$. This means that the outage probability for user $n$ is no longer a function $a_n$ because $\mathcal{B}_n^\text{out} = \left\{|G_n|^2 < \tfrac{2^{R_m}-1}{\xi(a_m-(2^{R_m}-1)\sum_{l=m+1}^K a_l)} \right\}$ remains constant. Hence, power allocation to user $n$ can essentially be increased without any benefit to performance, which is something that should be avoided. 
\subsection{NOMA power allocation strategies that achieve outage performance equal to OMA} 
Another requirement for a NOMA power allocation strategy is that the outage probability performance is equal to or better than the outage probability performance of OMA. 
First, the following power allocation strategies are formally defined in the following.
\begin{definition}\label{def:NOMA_power_def}
	For a user $n$:
		\begin{enumerate}[(i)]
			\item The power allocation coefficient $a_n^\text{oma}$ is defined as the exact power allocation required such that user $n$ achieves the same outage probability performance as it would achieve using OMA. In other words, $\mathcal{B}_n = \mathcal{B}_n^\text{oma}$;
			\item The power allocation coefficient $\tilde{a}_n^\text{oma}$ is defined as the minimum power allocation such that $\mathcal{B}_n = \mathcal{B}_n^\text{oma}$, which can only be applied when all users $l=n+1,\ldots,K$ also have power allocation $\tilde{a}_l^\text{oma}$;
			\item The interference coefficient $A_n^\text{oma} = \sum_{l=n+1}^K \tilde{a}_l^\text{oma}$.
		\end{enumerate}
\end{definition} 

Any power allocation strategy that improves the outage probability performance over OMA can be written as $(a_1^\text{oma}+\epsilon_1,\ldots,a_K^\text{oma}+\epsilon_K)$. If $\epsilon_n=0, \forall n=1,\ldots,K$, then all users will achieve the same outage probability performance as OMA, and $a_n^\text{oma} = \tilde{a}_n^\text{oma}, \forall n=1,\ldots,K$. 
The following theorem shows that there always exists a power allocation strategy such that the NOMA outage probabilities for all users are equal to or less than the respective OMA outage probabilities. 

\begin{theorem}\label{thm:NOMA_OMA_strategy}
	For a $K$-user DL NOMA system with target rates $R_1,\ldots,R_K$, there always exists a power allocation strategy $(a_1,\ldots,a_K)$ with associated SIC decoding order $(1,\ldots,K)$ such that $\mathcal{B}_m\subseteq\mathcal{B}_m^\text{oma},\forall n=1,\ldots,K$. Furthermore, $\exists$ at least one user $n$ such that $\mathcal{B}_n\subset\mathcal{B}_n^\text{oma}$, meaning that the NOMA outage probability performance can always be at least as good or better than the OMA outage probability performance for every user.
\end{theorem}
\begin{IEEEproof}
	See appendix \ref{proof:NOMA_OMA_strategy}.
\end{IEEEproof}

According to equation (\ref{eq:NOMA_OMA_equality}), $(\tilde{a}_1^\text{oma},\ldots, \tilde{a}_K^\text{oma})$ is given by $\tilde{a}_K^\text{oma}  = \frac{2^{R_K}-1}{2^{R_K/\tau_K}-1}$ and
\begin{equation*}
	\tilde{a}_n^\text{oma} = \tfrac{2^{R_n}-1}{2^{R_n/\tau_n}-1} + \tfrac{2^{R_n}-1}{2^{R_n}}\sum_{l=n+1}^K\tfrac{2^{R_l}-1}{2^{R_l/\tau_l}-1}\prod_{k=n}^{l-1}2^{R_k}, 
\end{equation*}
$n=1,\ldots,K-1$. Based on the previous theorem, it is clear that $\sum_{n=1}^K\tilde{a}_n^\text{oma}<1$, and that the improvement of the outage probability performance of NOMA over OMA is based on the design of the additional power allocation $\epsilon_n$ to each coefficient $a_n^\text{oma}$, and the strategy $(\tilde{a}_1^\text{oma},\ldots, \tilde{a}_K^\text{oma})$ is the starting point. A consequence of theorem \ref{thm:NOMA_OMA_strategy} is that it can be used to highlight the fact that the decoding order based on increasing values of $\frac{R_{n}}{\tau_n}$ is an essential component of the power allocation strategy.
\begin{corollary}\label{cor:EE-decode-order}
Let the user indices $1,\ldots, K$ be assigned such that they follow the relationship $\frac{R_{1}}{\tau_1}<\cdots<\frac{R_{K}}{\tau_K}$. Also, define a SIC decoding order $(\sigma(1), \ldots, \sigma(K))$, such that $(\sigma(1), \ldots, \sigma(K))$ is a permutation of the sequence $(1,\ldots,K)$. 
For all SIC decoding orders $(\sigma(1), \ldots, \sigma(K))$ which have associated power allocation strategies $(\tilde{a}_{\sigma(1)}^\text{oma},\ldots, \tilde{a}_{\sigma(K)}^\text{oma})$ such that NOMA achieves equal outage performance to OMA, the SIC decoding order
\begin{equation}
	(1,\ldots,K) = \arg\min_{(\sigma(1),\ldots,\sigma(K))}\sum_{n=1}^K \tilde{a}_{\sigma(n)}^\text{oma}.
\end{equation} 
In other words, the most energy efficient power allocation strategy is the one where $(\sigma(1), \ldots, \sigma(K)) = (1,\ldots,K)$. 
\end{corollary} 
\begin{IEEEproof}
	See appendix \ref{proof:EE-decode_order}.
\end{IEEEproof}
This corollary states that the most energy efficient power allocation strategy which enables NOMA outage performance equal to that of OMA is based on the SIC decoding order which follows the increasing order of $\frac{R_n}{\tau_n}$. The most important aspect of this result is that this SIC decoding order provides the most power allocation headroom in order to improve the outage performance of NOMA over OMA.
In the case that user $m$ has power allocation greater than $\tilde{a}_m^\text{oma}$, then clearly all users $n=1, \ldots,m-1$ will have to allocate additional power in order for $\mathcal{B}_n=\mathcal{B}_n^\text{oma}$. Furthermore, any power allocation strategy should be demonstrated to be well-behaved. The fundamental properties of well-behaved NOMA power allocation strategies which demonstrates better outage probability performance over OMA are discussed in the next section.

\section{Well-behaved power allocation strategies that improve NOMA outage probability performance over OMA}\label{sec:NOMApowerstrat}
In order to determine how to construct a well-behaved power allocation strategy which improves NOMA outage probability performance over OMA, the power allocation  strategy that satisfies theorem \ref{thm:NOMA_OMA_strategy} must be generalized. Since a power allocation coefficient for user $n$'s signal can be described by $a_n = a_n^\text{oma}+\epsilon_n, \forall n$, $a_n^\text{oma} = \frac{2^{R_n}-1}{2^{KR_n}-1}+(2^{R_n}-1)A_n$ (where $A_n\geq A_n^\text{oma}$), and $a_K^\text{oma} = \tilde{a}_K^\text{oma}$, then 
\begin{align}
	&a_K = \tilde{a}_K^\text{oma}+\epsilon_K\\
	&a_{K-1} = a_{K-1}^\text{oma}+\epsilon_{K-1} = \tilde{a}_{K-1}^\text{oma}+\epsilon_{K-1}+(2^{R_{K-1}}-1)\epsilon_K \nonumber\\
	&a_{n} = \tilde{a}_{n}^\text{oma}+\epsilon_n+(2^{R_{n}}-1)\left(\epsilon_{n+1}+\sum_{l=n+2}^K\epsilon_l\prod_{k=n+1}^{l-1}2^{R_k}\right), \nonumber
\end{align}
for $n=1,\ldots,K-2$. Note that by definition \ref{def:NOMA_power_def}, $a_n=a_n^\text{oma}$ iff $\epsilon_n=0$, and $a_n^\text{oma} = \tilde{a}_n^\text{oma}$ iff $\epsilon_l=0, \forall l=n+1,\ldots,K$. Furthermore, the portion of the interference coefficient caused by the terms $\epsilon_l, l=n+1,\ldots,K$ (the expression in the parenthesis above) can be expressed as 
\begin{align}
	&c_n = \epsilon_{n+1}+\sum_{l=n+2}^K\epsilon_l\prod_{k=n+1}^{l-1}2^{R_k},
\end{align} 
So the general interference coefficient for user $n$ can be written as $A_n = A_n^\text{oma}+c_n$. 

The total available power allocation coefficient for user $n$ is a function of $\epsilon_m, m=1,\ldots, n-1$. This is because in a DL NOMA system, the goal is to improve the overall outage performance, and the outage performance of the users later in the SIC decoding order is more difficult to improve, as shown by the coefficient $c_n$. Thus, improving the performance of users with signals earlier in the SIC decoding order does not come at an additional cost for users later in the decoding order. The total available power allocation coefficient for user $n$ is then found by noting that
\begin{align}
	A_\text{tot}^\text{oma} + \epsilon_1 + \sum_{m=1}^{n-1}\epsilon_m\prod_{k=1}^{m-1}2^{R_k} < 1.
\end{align}
The sum of the additional power allocation for users $m=1, \ldots,n-1$, is given by
\begin{equation}
	d_n = \epsilon_1 + \sum_{l=2}^{n-1}\epsilon_l\prod_{k=1}^{l-1}2^{R_k}.
\end{equation}
So the additional power allocation coefficient $\epsilon_n$ for user $n$ is a function of $d_n$.

Using the generalized expression of the power allocation strategy that satisfies theorem \ref{thm:NOMA_OMA_strategy}, the properties of $\epsilon_n$ can be found such that the power allocation strategy is well-behaved.
\begin{theorem}\label{thm:well-behaved-NOMA}
	For users $1,\ldots,K$ with target rates $R_1, \ldots, R_K$, which are scheduled to receive signals with power allocation strategy $(a_1^\text{oma}+\epsilon_1,\ldots,a_K^\text{oma}+\epsilon_K)$, the power allocation strategy is well-behaved if each user $n$ has one or the other of the following conditions:
	\begin{enumerate}[(a)]
		\item $a_{n-1} = a_{n-1}^\text{oma}$ and $a_{n} = a_{n}^\text{oma}$, meaning $\epsilon_{n-1}$ and $\epsilon_n=0$, for any $n=2,\ldots,K$;
		\item $0<\epsilon_{n}$
	\end{enumerate}
			\begin{align}
				\hspace{3mm}\label{eq:epsilon_bound}<\min\left\{ \begin{array}{l} \epsilon_{n-1}\frac{2^{R_{n}}-1}{2^{R_{n-1}}-1}+\frac{2^{R_{n}}-1}{2^{R_{n-1}/\tau_{n-1}}-1} - \frac{2^{R_{n}}-1}{2^{R_{n}/\tau_n}-1},  \\ 
									\displaystyle(1-A_\text{tot}^\text{oma})\prod_{l=1}^{n-1}2^{-R_l} - \sum_{m=1}^{n-1}\epsilon_m\prod_{l=l}^{n-1}2^{R_l} \end{array}\right\}.
			\end{align}
\end{theorem}
\begin{IEEEproof}
	See appendix \ref{proof:well-behaved-NOMA}.
\end{IEEEproof}

Now that the fundamental properties of well-behaved additional power allocation strategies beyond $\tilde{a}_n^\text{oma}$ have been described in detail, the method for selecting/designing a power allocation strategy can be discussed. Specifically, the selection/design of the power allocation strategy is completely focused on the selection of $\epsilon_n,n=1,\ldots,K$. In other words, if an algorithm is designed to minimize the overall outage probability performance with respect to the power allocation strategy, and subject to the constraints that the performance of each user outage is better than the OMA performance, then the variables to be solved for are $(\epsilon_1,\ldots,\epsilon_K)$, and the constraints are given by theorem \ref{thm:well-behaved-NOMA}. These constraints are linear with coefficients based on the target rates $R_1,\ldots,R_K$ and OMA time durations $(\tau_1,\ldots,\tau_K)$. 

However, a simpler but not optimal approach can be used to determine a power allocation strategy such that it satisfies theorem \ref{thm:well-behaved-NOMA} by using the definition of being {\em well-behaved}, hence enhancing the outage probability performance of each user with respect to OMA. This is accomplished by noting that if
\begin{align}
	\label{eq:add_well_behaved}&\epsilon_{n-1}>\epsilon_n2^{R_n}\tfrac{2^{R_{n-1}}-1}{2^{R_n}-1}, n=2,\ldots,K
\end{align}
then the power allocation strategy is well-behaved. This can be accomplished using the total addition power allocated to all users $m = 1,\ldots,n-1$ caused by adding $\epsilon_n$ to user $n$'s power allocation. When user $n$ has $\epsilon_n$ added to its power allocation coefficient, the BS must also add to the power allocation coefficient of users $m=1,\ldots,n-1$ in order to maintain their outage performance, according to equations (\ref{eq:add_power_compensate1}) and (\ref{eq:add_power_compensate2}).
This amount can be easily seen to be
\begin{equation}
	\epsilon_n^\text{tot} = \epsilon_n\prod_{l=1}^{n-1}2^{R_l}.
\end{equation} 
Therefore, by setting
\begin{align}
	\label{eq:total_add_power_well_behaved}&\epsilon_{n-1}^\text{tot} = \epsilon_n^\text{tot}2^{R_n}\tfrac{2^{R_{n-1}}-1}{2^{R_n}-1}, n=2,\ldots,K
\end{align}
it can easily be shown that equation (\ref{eq:add_well_behaved}) is satisfied. Let $S = \sum_{n=1}^K R_n$. Solving the $K-1$ equations in equation (\ref{eq:total_add_power_well_behaved}) for $\epsilon_K$, and using the fact that the sum of the additional power allocation coefficients is bounded by $1-A_\text{tot}^\text{oma}$, yields 
\begin{align}
	\label{eq:well-behaved-strategy}&\sum_{n=1}^K\epsilon_n^\text{tot} = 1-A_\text{tot}^\text{oma}\\
	\Rightarrow&\epsilon_K = (1-A_\text{tot}^\text{oma})\frac{2^{R_K}-1}{2^S-1}\prod_{l=1}^{K-1}2^{-R_l}\nonumber \\
	&\epsilon_1 = (1-A_\text{tot}^\text{oma})\frac{2^{R_1}-1}{2^S-1}\prod_{l=2}^K 2^{R_l},\nonumber\\
	&\epsilon_n = (1-A_\text{tot}^\text{oma})\frac{2^{R_n}-1}{2^S-1}\prod_{l=1}^{n-1}2^{-R_l}\prod_{l=n+1}^K 2^{R_l},\nonumber
\end{align}
for $n=2,\ldots,K-1$.

This simple strategy will use all of the power allocation available, while improving the outage probability performance of all $K$ users when employing NOMA over OMA. Note that this strategy also heavily distributes the remaining available power allocation coefficient $1-A_\text{tot}^\text{oma}$ in favor of the users whose signals are earlier in the SIC decoding order. This is in line with what is expected with DL NOMA systems with SIC enabled receivers, where users whose signals are decoded first will have their interference removed, and thus the additional power allocation coefficient $\epsilon_n$ will 
also improve the SIC performance of users $l=n+1,\ldots,K$. While users whose signals are later in the SIC decoding order cause interference which in turn causes all users $m=1,\ldots,n$ to have their power allocation coefficient bumped up in order to maintain the same performance, and thus creating the case where less additional power allocation is actually available and gains are marginal.

\section{Comparison of theoretical and simulation results} \label{sec:results}

For the simulation results, two different fading channel scenarios are used to demonstrate the validity of the theoretical results. The first fading channel model is the $K$ SISO Rayleigh fading channel, with channel gains $h_{n_1},\ldots,h_{n_K}$, such that $|h_{n_i}|\sim\mathrm{Exponential}(1), i=1,\ldots,K$. The channel SNR gains $|G_1|^2,\ldots,|G_K|^2 = \mathrm{sort}(|h_{n_1}|^2,\ldots,|h_{n_K}|^2)$, where the $\mathrm{sort}$ function sorts the channel SNR gains in ascending order. Therefore, $|G_1|^2<\ldots<|G_K|^2$. This is conceptually the same model used in \cite{NOMA-RDU:DYFP,5GNOMA:DFP,FairNOMAFull}, where the ordering of i.i.d. Rayleigh fading channel gains are used to represent the position of a user within a cell, and thus outage probabilities and diversity orders are derived from the distribution of this ordering. For the simulations using this channel model, the ordering of the channel gains and that of the SIC decoding order follow the same trend, so the user with weakest channel has its signal decoded first by all users, then the second weakest user, and so on. This channel model from here on is referred to as channel model 1.

The second channel model used is the MIMO Rayleigh fading channel model with i.i.d. fading channel gains between the different transmit-receive antenna pairs. A common precoding vector $\boldsymbol{p}, \|\boldsymbol{p}\|=1$, is used to transmit to $K$ users using $M$ antennas, where $\boldsymbol{p}$ is not a function of the channel gains\footnote{In cellular deployments, the precoder is typically selected from a set of predetermined vectors, based on CSI feedback}. The signal passes through user $n$'s  $N\times M$ channel matrix $\boldsymbol{H}_n$ where the channel from transmit antenna $i$ to receive antenna $j$ is $h_{j,i}\sim\mathcal{CN}(0,\beta_n)$, and each user $n$ with $N$ receive antennas uses the optimum detection vector $\boldsymbol{v}_n = \boldsymbol{p}^\mathsf{H}\boldsymbol{H}_n^\mathsf{H}/\|\boldsymbol{H}_n\boldsymbol{p}\|$. This gives a channel SNR gain of $|G_n|^2 = \|\boldsymbol{H}_n\boldsymbol{p}\|^2$. The channel SNR gain has distribution $\mathrm{Erlang}(\beta_n, N)$, with expected value $\mathbb{E}\left[|G_n|^2 \right] = N\beta_n$. For the simulations that use this channel model, $M=2$, $N=3$, $\boldsymbol{p}$ is selected randomly isotropically, and the $\beta_n$ are not selected with any relationship to the target rates in order to demonstrate that the channel gain ordering has no bearing on the validity of the results. Therefore, $(\beta_1=0.5, \beta_2=1.4, \beta_3=0.8, \beta_4=1.7, \beta_5=1.1)$. This channel model from here on is referred to as channel model 2.

For all simulation plots, there are $K=5$ users, the target rates are $(R_1=0.5,\allowbreak R_2=1.2,\allowbreak R_3=0.9,\allowbreak R_4=1.3,\allowbreak R_5=1.1)$ bps/Hz, and the OMA time durations are $(\tau_1=0.15, \tau_2=0.30, \tau_3=0.20, \tau_4=0.20, \tau_5=0.15)$. As mentioned previously, the decoding order must be such that $r_n=\frac{R_n}{\tau_n}$ is increasing, so  since $(r_1=\frac{10}{3}, r_2=4, r_3=4.5, r_4=6.5, r_5=\frac{22}{3})$, the indices for the rates and time durations above are as such.

\begin{figure}
	\centering
	\includegraphics[scale=0.4]{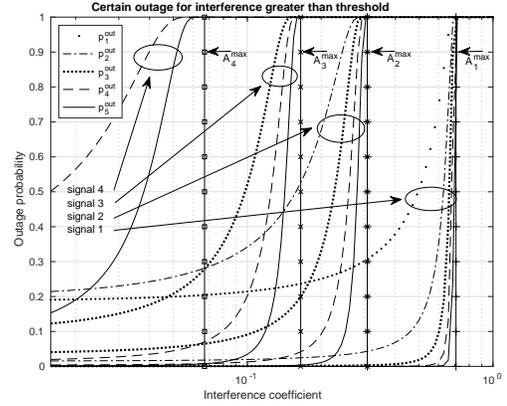} 
	\caption{\label{fig:sureout_ordered} {\bf Channel model 1}: Certain outage when interference exceeds limits in theorem \ref{thm:power_limit}; $\xi=10$dB} 
\end{figure}

\begin{figure}
	\centering
	\includegraphics[scale=0.4]{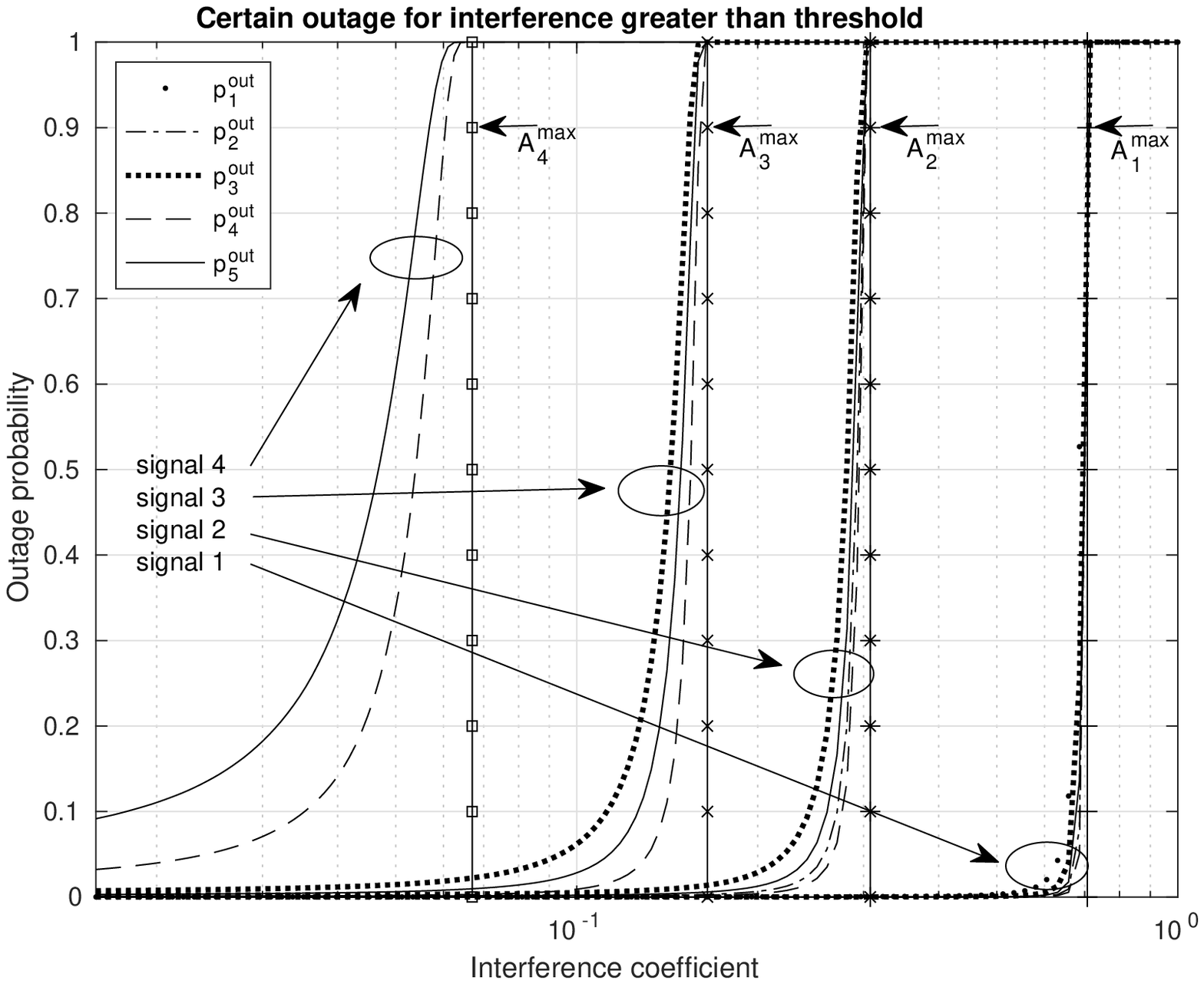} 
	\caption{\label{fig:sureout_statistical} {\bf Channel model 2}: Certain outage when interference exceeds limits in theorem \ref{thm:power_limit}; $\xi=10$dB} 
\end{figure}

Figure \ref{fig:sureout_ordered} and figure \ref{fig:sureout_statistical} demonstrate the phenomenon described in theorem \ref{thm:power_limit}. For a power allocation strategy such that the interference coefficient $A_n$ received when attempting to decode signal $x_n$ exceeds the value given in theorem \ref{thm:power_limit}, then the outage probability is equal to 1, regardless of the channel strength and SNR. As can be seen in figure \ref{fig:sureout_ordered}, for each signal to be decoded, the outage probabilities are lesser for users with stronger channels. In figure \ref{fig:sureout_statistical}, the same phenomenon is observed even though the users have more receive antennas to increase their received SNR. In this case user 4 has the strongest channel statistically, so user 4 always has the least outage probabilities when the interference is below the certain outage threshold. 
 
\begin{figure}
	\centering
	\includegraphics[scale=0.4]{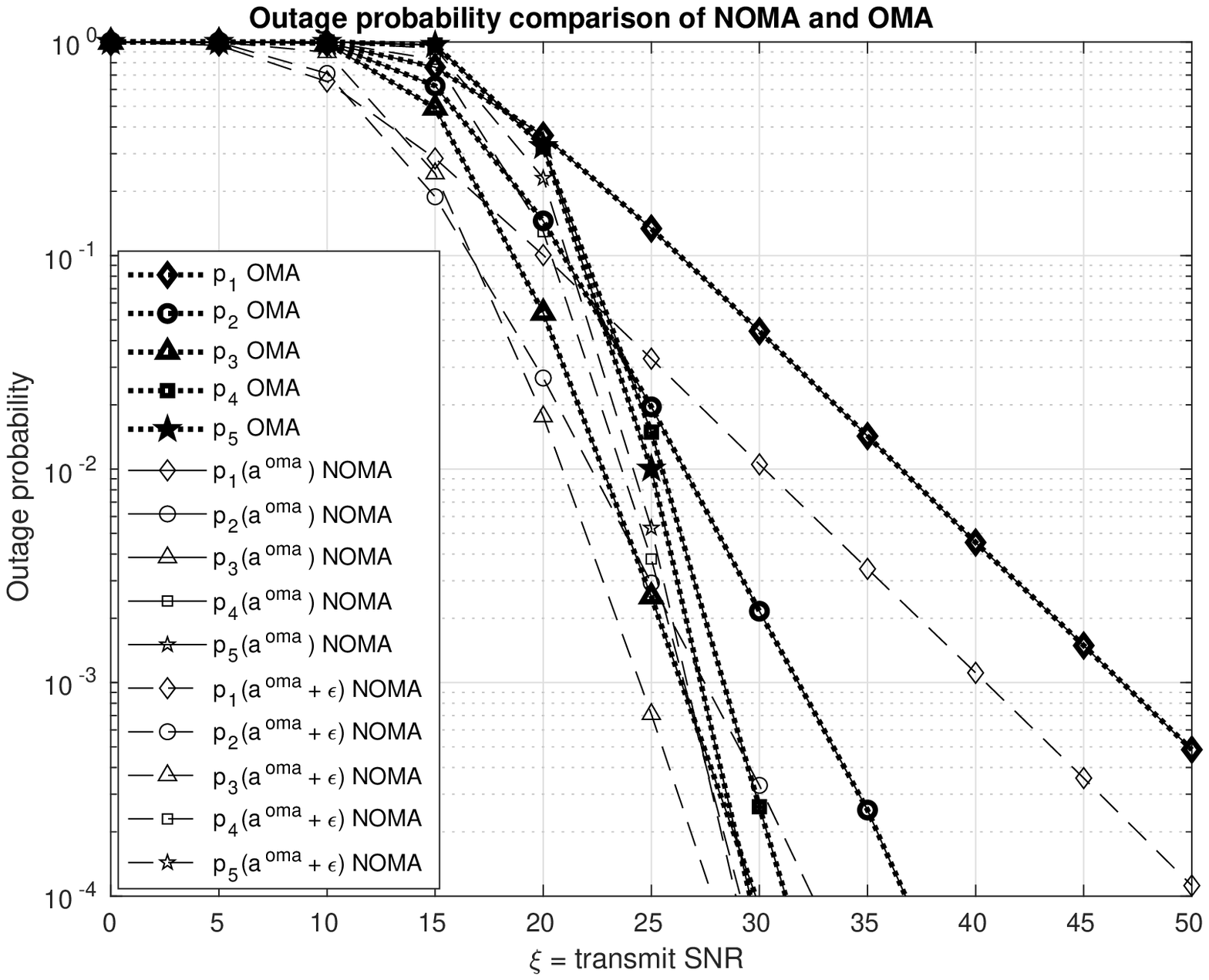} 
	\caption{\label{fig:omanoma_ordered} {\bf Channel model 1}: Comparing NOMA performance according to power allocation strategies that perform equal to OMA (theorem \ref{thm:NOMA_OMA_strategy}) and better than OMA} 
\end{figure}

\begin{figure}
	\centering
	\includegraphics[scale=0.4]{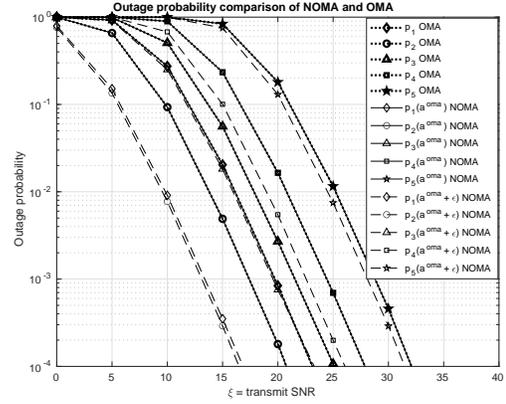} 
	\caption{\label{fig:omanoma_statistical} {\bf Channel model 2}: Comparing NOMA performance according to power allocation strategies that perform equal to OMA (theorem \ref{thm:NOMA_OMA_strategy}) and better than OMA} 
\end{figure}

Figures \ref{fig:omanoma_ordered} and \ref{fig:omanoma_statistical} demonstrate the outage probability performance for NOMA compared to OMA, when NOMA uses both the power allocation strategy $(\tilde{a}_1^\text{oma},\ldots,\tilde{a}_5^\text{oma})$ in order to demonstrate the validity of the power allocation result of theorem \ref{thm:NOMA_OMA_strategy} (blue curves), and $(a_1^\text{oma}+\epsilon_1,\ldots,a_5^\text{oma}+\epsilon_5)$ with $\epsilon_n$ selected according to the power allocation strategy described in equation (\ref{eq:well-behaved-strategy}) to ensure the power allocation coefficients are well-behaved (red curves). Clearly when the power allocation strategy for NOMA is $(\tilde{a}_1^\text{oma},\ldots,\tilde{a}_5^\text{oma})$, the outage probability is exactly equal to that of OMA. However,  as proven in theorem \ref{thm:NOMA_OMA_strategy} the sum of the power allocation is less than 1. In fact, for this particular case it is $\approx 0.5036$, which means that roughly only half of the maximum transmit SNR is needed to have the outage performance of NOMA equal that of OMA. In figure \ref{fig:omanoma_ordered}, there is a large difference in performance between NOMA (red curves) and OMA outage probabilities for users $K=1,2,3$, while for users $K=4,5$ the gap is not so big. The same phenomenon is observed in figure \ref{fig:omanoma_statistical}, even though the ordering of the users' channel gains is not considered in the SIC decoding order. It makes sense that the gap in outage probability performance decreases for users whose signals are decoded towards the end of the SIC procedure. For example, if the BS tries to improve user 5's outage probability performance using NOMA over OMA by allocating $\epsilon_5$ additional power allocation coefficient to its signal, while keeping the outage performance of the other users the same as OMA, the BS also has to increase the power allocation coefficient of user 4 by $c_4=\epsilon_5(2^{R_4}-1)$, and for user $n$ by $c_n = \epsilon_5(2^{R_n}-1)\prod_{k=n+1}^{4}2^{R_k}, n=1,2,3$, just so that they can have the same performance as OMA. So the amount of additional power allocation that the BS has available for a signal that is decoded later in the SIC procedure becomes less. 

\begin{figure}
	\centering
	\includegraphics[scale=0.4]{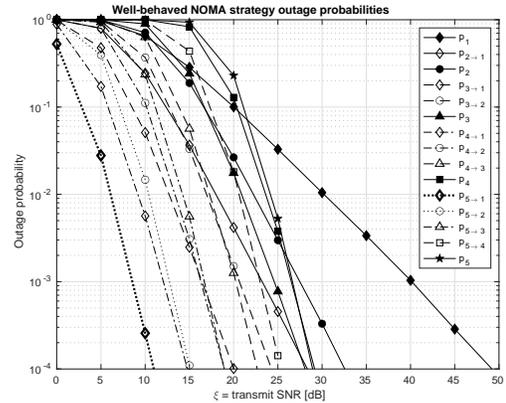} 
	\caption{\label{fig:well-behaved_ordered} {\bf Channel model 1}: Demonstration of well-behaved strategy behavior for NOMA}
\end{figure}

\begin{figure}
	\centering
	\includegraphics[scale=0.4]{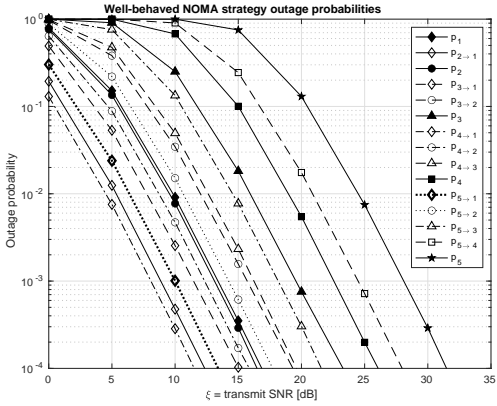} 
	\caption{\label{fig:well-behaved_statistical} {\bf Channel model 2}: Demonstration of well-behaved strategy behavior for NOMA} 
\end{figure}

In figures \ref{fig:well-behaved_ordered} and \ref{fig:well-behaved_statistical}, the well-behaved property of the strategy derived is demonstrated by plotting the outage probabilities for each signal to be decoded by each user in the SIC procedure. For example, user 5 must decode signals 1, 2, 3, and 4 before it can decode its own signal, and the outage probability performances are better for the signals earlier in the SIC procedure. A similar phenomenon is observed for user 4 with signals 1, 2, and 3, and so for the other users. This is consistent with what was stated regarding the overall outage event for a user, as it should not be bounded by the outage event of an earlier signal in the SIC procedure. In other words, the probability of outage should always be better for the decoding of the signals that are earlier in the SIC procedure. 
For figure \ref{fig:well-behaved_ordered}, the outage probability for decoding a specific signal, say signal 1, is better for the users with stronger channels, as can be seen by the blue diamond curve belonging to user 5 being the best for decoding signal 1, and the black diamond curve belonging to user 1 being the worst, which is still better than user 1's outage probability curve for OMA as shown in \ref{fig:omanoma_ordered}. 
The same phenomenon is observed in figure \ref{fig:well-behaved_statistical}, except that here the user with statistically the strongest channel gain is user 4, and accordingly the red diamond curve belonging to user 4 outperforms all of the other diamond curves. In this plot, even though user 5 is only the third strongest channel out of all, it has the signal that is decoded last among all other signals, and thus decodes all four other signals first, yet its outage probabilities for the first four decodings still demonstrate a well-behaved power allocation, while the outage curve with the blue star is still better than its OMA outage curve given (both seen in figure \ref{fig:omanoma_statistical}).

\section{Conclusion and Future Work}\label{sec:conclusion}

In this work, it was demonstrated that for downlink NOMA systems with a BS which does not have knowledge of the exact channel gains, the power allocation strategy must be carefully designed in order to avoid certain outages for multiple users. Furthermore, it was demonstrated that a well-behaved power allocation strategy which has the same exact outage performance as OMA always exists, such that it depends only on the target rates and their relative OMA time durations, and it is derived in closed form. The proposed SIC decoding order based on increasing values of $\frac{R_n}{\tau_n}$ was shown to be the most energy efficient. Lastly, the approach for designing a power allocation strategy which can always outperform OMA in terms of outage probability was outlined, and the associated properties of such a strategy were derived. The validity of these theoretical results are then substantiated with the simulation results, which show the outage performances for various power allocation strategies to exhibit these fundamental characteristics outlined in the paper. 

One thing that is not addressed in this paper is the fact that the channel SNR gains can be used in the design of the power allocation coefficients. Comparing these results to the multi-user approaches similar to that in \cite{FairNOMAFull} can provide a very quick and simple assessment as to whether the channel SNR gains are strong enough to support the target rates. Further studies about how this type of phenomenon is exhibited and described theoretically in more complex cellular deployments is also critical, such as in multi-point and heterogeneous cellular networks. Lastly, a full treatment of the uplink scenario with regards to the power allocation strategy design is needed, as uplink NOMA is sought to be a vital deployment scenario for 5G cellular systems.


\appendices
\renewcommand{\theequation}{\thesection.\arabic{equation}}
\section{Proofs}
\subsection{Proof of Theorem \ref{thm:power_limit}}\label{proof:power_limit}
\begin{IEEEproof}
	For any specific user $n$, suppose that $A_{n-1} < 2^{-\sum_{l=1}^{n-1}R_l}$ and $A_n > 2^{-\sum_{l=1}^{n}R_l}$. Since $A_{n-1}=a_n+A_n$, it follows that 
	\begin{align}
		&a_n+A_n < 2^{-\sum_{l=1}^{n-1}R_l} \\ 
		&\Longrightarrow a_n < 2^{-\sum_{l=1}^n R_l}-A_n < 2^{-\sum_{l=1}^{n-1}R_l} - 2^{-\sum_{l=1}^{n}R_l}. \nonumber
	\end{align}
	The events $\mathcal{B}_n$ and $\mathcal{B}_{k\rightarrow n}$ can be written in the form
	\begin{align}
		&\log_2\left( 1 + \tfrac{a_n\xi |G_k|^2}{A_n\xi |G_k|^2+1}\right) < R_n, k=n,\ldots,K \nonumber\\
		\label{eq:thm1-inq1}&\Longrightarrow\xi |G_k|^2 ( a_n -(2^{R_n}-1)A_n) < 2^{R_n}-1.
	\end{align}
	Since $a_n<2^{-\sum_{l=1}^{n-1}R_l} - 2^{-\sum_{l=1}^n R_l}$ and $A_n>2^{-\sum_{l=1}^{n}R_l}$, 
	\begin{align}
		&a_n -(2^{R_n}-1)A_n \nonumber\\
		&<2^{-\sum_{l=1}^{n-1}R_l} - 2^{-\sum_{l=1}^{n}R_l} -(2^{R_n}-1)A_n \nonumber\\
		&<2^{-\sum_{l=1}^{n-1}R_l} - 2^{-\sum_{l=1}^{n}R_l} -(2^{R_n}-1)2^{-\sum_{l=1}^{n}R_l} \nonumber\\
		\label{eq:thm1_LB}&=0.
	\end{align}
	Therefore solving equation (\ref{eq:thm1-inq1}) for $|G_k|^2$ leads to
	\begin{align}
		&\xi |G_k|^2 ( a_n -(2^{R_n}-1)A_n) < 2^{R_n}-1\\ 
		\label{eq:certain_outage}&\Longrightarrow |G_k|^2 > \tfrac{2^{R_n}-1}{\xi(a_n-(2^{R_n}-1)A_n)}.\nonumber
	\end{align}
	Therefore, since $\frac{2^{R_n}-1}{\xi(a_n-(2^{R_n}-1)A_n)}<0<|G_k|^2$, this condition makes $\mathrm{Pr}\{\mathcal{B}_n\} = \mathrm{Pr}\{\mathcal{B}_{k\rightarrow n}\}=1$. 
	
	Now suppose that $A_n > 2^{-\sum_{l=n+1}^{K}R_l}, \forall n=1,\ldots,K-1$, then it must be true that $A_1 > 2^{-R_1}$. This will avoid the previous impossible event. However, if this is true, then events $\mathcal{B}_1$ and $\mathcal{B}_{k\rightarrow 1}, k=2,\ldots,K,$ gives rise to the inequality
	\begin{align}
		&\xi |G_k|^2 (a_1 -(2^{R_1}-1)A_1) < 2^{R_1}-1, k=1,\ldots,K,
	\end{align}
	where the value inside the parentheses must be greater than zero in order to avoid the certain outage situation from equation (\ref{eq:certain_outage}). Therefore, 
	\begin{align}
		&0< a_1 -(2^{R_1}-1)A_1 < a_1 -(2^{R_1}-1)2^{-R_1} \nonumber \\
		\Rightarrow& a_1 > 1-2^{-R_1}. 
	\end{align}
	It must be true that $a_1+A_1\leq 1$ by definition of power allocation coefficients, however
	\begin{align}
		&a_1+A_1 > (1-2^{-R_1}) + 2^{-R_1} = 1.
	\end{align}
	Therefore if $A_n>2^{-\sum_{l=1}^{n}R_l}, \forall n=1,\ldots,K-1$, then having $\mathrm{Pr}\{\mathcal{B}_n\} <1$ and $\mathrm{Pr}\{\mathcal{B}_{k\rightarrow n}\}<1$ requires $\sum_{n=1}^K a_n = a_1+A_1 > 1$, which is not possible. 
	
	Hence, for any user $n$ with $A_n>2^{-\sum_{l=1}^{n}R_l}$, $\mathrm{Pr}\{\mathcal{B}_n\} = \mathrm{Pr}\{\mathcal{B}_{k\rightarrow n}\} = 1, k=n+1,\ldots,K$.	
\end{IEEEproof}

\subsection{Proof of Proposition \ref{prop:power_coeff_relate}}\label{proof:power_coeff_relate}
\begin{IEEEproof}
	If $\mathcal{B}_n^\text{out} = \mathcal{B}_n, \forall n=2,\ldots,K$, then it is true that 
	\begin{align}
		&\frac{2^{R_{m}}-1}{a_{m}-(2^{R_{m}}-1)\sum_{l=m+1}^K a_l}\leq\frac{2^{R_{n}}-1}{a_{n}-(2^{R_{n}}-1)\sum_{l=n+1}^K a_l},
	\end{align}
	$\forall m=1,\ldots,n-1$. If this is true, then it is true that
	\begin{align}
		&\frac{2^{R_{1}}-1}{a_{1}-(2^{R_{1}}-1)\sum_{l=2}^K a_l}\leq\cdots\\
		&\leq \frac{2^{R_{K-1}}-1}{a_{K-1}-(2^{R_{K-1}}-1)\sum_{l=K}^K a_l}\leq \frac{2^{R_{K}}-1}{a_K}. \nonumber
	\end{align}
	From the above, it is easy to show that 
	\begin{equation}\label{eq:power_coeff_bound}
		a_{n-1}\geq a_n \frac{2^{R_n}(2^{R_{n-1}}-1)}{2^{R_n}-1}, n = 2,\ldots, K.
	\end{equation}
	To show that the condition above implies that $a_1,\ldots,a_K$ satisfy theorem \ref{thm:power_limit}, it is sufficient to show that any power allocation coefficients satisfying equation (\ref{eq:power_coeff_bound}) satisfy the inequality 
	\begin{equation}
		a_n-(2^{R_n}-1)A_n>0,
	\end{equation}
	based on equation (\ref{eq:thm1_LB}), according to the theorem. So if equation (\ref{eq:power_coeff_bound}) holds $\forall n=2,\ldots,K$, then for any $n<K$ and $l>n$, it is easily shown that 
	\begin{align}
		&a_l< a_n\frac{2^{R_l}-1}{(2^{R_n}-1)\prod_{m=n+1}^l 2^{R_m}}\nonumber\\
		&\Longrightarrow \sum_{l=n+1}^K a_l < \frac{a_n}{2^{R_n}-1}\sum_{l=n+1}^K\frac{2^{R_l}-1}{\prod_{m=n+1}^l 2^{R_m}}\nonumber\\
		&= \frac{a_n}{2^{R_n}-1}\left( 1+ \sum_{l=n+2}^K\prod_{m=n+1}^{l-1}\frac{1}{2^{R_m}} -\sum_{l=n+1}^K\prod_{m=n+1}^l\frac{1}{2^{R_m}}\right)\nonumber\\
		&= \frac{a_n}{2^{R_n}-1}\left( 1 - \prod_{m=n+1}^K 2^{-R_m}\right).
	\end{align}
	So
	\begin{align}
		&a_n-(2^{R_n}-1)\sum_{l=n+1}^K a_l \nonumber \\
		&> a_n-(2^{R_n}-1)\frac{a_n}{2^{R_n}-1}\left( 1 - \prod_{m=n+1}^K 2^{-R_m}\right)\nonumber\\
		&= a_n\prod_{m=n+1}^K 2^{-R_m}>0, \forall n=1,\ldots,K-1.
	\end{align}
	Hence, these power allocation coefficients satisfy the requirement in theorem \ref{thm:power_limit}.
\end{IEEEproof}

\subsection{Proof of Theorem \ref{thm:NOMA_OMA_strategy}}\label{proof:NOMA_OMA_strategy}
\begin{IEEEproof}
	If $\mathcal{B}_n\subseteq\mathcal{B}_n^\text{oma}, \forall n=1,\ldots,K$, it must at least be true that $\exists (a_1,\ldots,a_K)$ s.t. $\mathcal{B}_n=\mathcal{B}_n^\text{oma}, \forall n=1,\ldots,K$, and then demonstrate that $\sum_{n=1}^K a_n < 1$. To show that $\exists (a_1,\ldots,a_K)$ s.t. $\mathcal{B}_n=\mathcal{B}_n^\text{oma}, \forall n=1,\ldots,K$, begin with $n=K$ and equate 
	\begin{align}
		&\frac{2^{R_K}-1}{a_K\xi} = \frac{2^{R_K/\tau_K}-1}{\xi}\label{eq:OMA_powercoeff_K} \hspace{3mm}\Rightarrow\hspace{3mm} a_K = \frac{2^{R_K}-1}{2^{R_K/\tau_K}-1}.
	\end{align}
	Then for $n=1,\ldots,K-1$, equate
	\begin{align}
		&\frac{2^{R_n}-1}{\xi(a_n-(2^{R_n}-1)\sum_{l=n+1}^K a_l)}=\frac{2^{R_n/\tau_n}-1}{\xi}\nonumber\\
		\label{eq:NOMA_OMA_equality}\Rightarrow& a_n-(2^{R_n}-1)\sum_{l=n+1}^K a_l = \frac{2^{R_n}-1}{2^{R_n/\tau_n}-1}
	\end{align}
	This creates a recursive relationship which can be solved to find
	\begin{align}
		a_n &= \frac{2^{R_n}-1}{2^{R_n/\tau_n}-1} + \frac{2^{R_n}-1}{2^{R_n}}\sum_{l=n+1}^K\frac{2^{R_l}-1}{2^{R_l/\tau_l}-1}\prod^{l-1}_{k=n}2^{R_k},\label{eq:OMA_powercoeff_n}
	\end{align}
	$n = 1,\ldots,K-1$. From here on, the power allocation strategy that satisfies equations (\ref{eq:OMA_powercoeff_K}, \ref{eq:OMA_powercoeff_n}) will be called $(\tilde{a}_1^\text{oma},\ldots,\tilde{a}_K^\text{oma})$.
	In order for this to be a valid power allocation strategy, the sum of the coefficients must be proven to always be less than or equal to 1. Let the interference coefficient for user $n$ using this power allocation strategy be called $A_n^\text{oma}$, which can be found easily by noting from equation (\ref{eq:NOMA_OMA_equality}) that
	\begin{align}
		\label{eq:power_oma_equal}&\tilde{a}_n^\text{oma} - (2^{R_n}-1)A_n^\text{oma} = \frac{2^{R_n}-1}{2^{R_n/\tau_n}-1}\\
		&\Rightarrow A_n^\text{oma} = \frac{1}{2^{R_n}-1}\left(\tilde{a}_n^\text{oma} - \frac{2^{R_n}-1}{2^{R_n/\tau_n}-1}\right)\nonumber\\
		\label{eq:NOMA_OMA_int}& = \frac{2^{R_{n+1}}-1}{2^{R_{n+1}/\tau_{n+1}}-1} + \sum_{l=n+2}^K\frac{2^{R_l}-1}{2^{R_l/\tau_l}-1}\prod_{k=n+1}^{l-1}2^{R_k}.
	\end{align}
	Define $\tau_n = \frac{b_n}{K}$ and $r_n = \frac{R_n}{b_n}$, so that $r_1<\cdots<r_K$. Since the function $h(t) = (2^{bt}-1)/(2^{Kt}-1)$ is a monotonically decreasing function in $t$ so long as $b<K$, then 
	\begin{align*}
		&A_n^\text{oma} = \frac{2^{R_{n+1}}-1}{2^{R_{n+1}/\tau_{n+1}}-1} + \sum_{l=n+2}^K\frac{2^{R_l}-1}{2^{R_l/\tau_l}-1}\prod_{k=n+1}^{l-1}2^{R_k}\\
		&=\frac{2^{b_{n+1}r_{n+1}}-1}{2^{Kr_{n+1}}-1} + \sum_{l=n+2}^K\tfrac{2^{b_lr_l}-1}{2^{Kr_l}-1}\prod_{k=n+1}^{l-1}2^{b_kr_k}\\
		& < \frac{2^{b_{n+1}r_{n+1}}-1}{2^{Kr_{n+1}}-1} + \sum_{l=n+2}^{K-2}\frac{2^{b_lr_l}-1}{2^{Kr_l}-1}\prod_{k=n+1}^{l-1}2^{b_kr_k} \\
		&+ \frac{2^{b_{K-1}r_{K-1}}-1}{2^{Kr_{K-1}}-1}\prod_{k=n+1}^{K-2}2^{b_kr_k} \\
		&+ \frac{(2^{b_Kr_{K-1}}-1)2^{b_{K-1}r_{K-1}}}{2^{Kr_{K-1}}-1}\prod_{k=n+1}^{K-2}2^{b_kr_k}\\
		& = \frac{2^{b_{n+1}r_{n+1}}-1}{2^{Kr_{n+1}}-1} + \sum_{l=n+2}^{K-2}\frac{2^{b_lr_l}-1}{2^{Kr_l}-1}\prod_{k=n+1}^{l-1}2^{b_kr_k} \\
		&+ \frac{2^{(b_{K-1}+b_K)r_{K-1}}-1}{2^{Kr_{K-1}}-1}\prod_{k=n+1}^{K-2}2^{b_kr_k}\\
		& < \cdots< \frac{2^{(b_{n+1}+\cdots+b_K)r_{n+1}}-1}{2^{Kr_{n+1}}-1}. 
	\end{align*}
	So given that $\tilde{a}_n^\text{oma} = \frac{2^{b_nr_n}-1}{2^{Kr_n}-1} + (2^{b_nr_n}-1)A_n^\text{oma}$, then
	\begin{align*}
		&A_\text{tot}^\text{oma}=\sum_{n=1}^K \tilde{a}_l^\text{oma} = \tilde{a}_1^\text{oma}+A_1^\text{oma}\\
		& = \frac{2^{b_1r_1}-1}{2^{Kr_1}-1} + (2^{b_1r_1}-1)A_1^\text{oma} + A_1^\text{oma}\\
		& = \frac{2^{b_1r_1}-1}{2^{Kr_1}-1} + 2^{b_1r_1}A_1^\text{oma}\\
		& < \frac{2^{b_1r_1}-1}{2^{Kr_1}-1} + 2^{b_1r_1}\frac{2^{(b_2+\cdots+b_K)r_{2}}-1}{2^{Kr_2}-1}\\
		& < \frac{2^{b_1r_1}-1}{2^{Kr_1}-1} + 2^{b_1r_1}\frac{2^{(b_2+\cdots+b_K)r_{1}}-1}{2^{Kr_1}-1}\\
		& = \frac{2^{(b_1+\cdots+b_K)r_1}-1}{2^{Kr_1}-1} = 1.
	\end{align*}
	So clearly the sum is less than 1. To complete the proof, only one strategy that satisfies the conditions stated in the theorem is needed, so let user $1$ have the power allocation coefficient $a_1 = a_1^\text{oma}+\epsilon_1$, such that
	\begin{align}
		&\epsilon_1 = 1 - A_\text{tot}^\text{oma} > 0,
	\end{align}
	which leads to $\mathcal{B}_1\subset\mathcal{B}_1^\text{oma}$.
\end{IEEEproof}

\subsection{Proof of Corollary \ref{cor:EE-decode-order}}\label{proof:EE-decode_order}
\begin{IEEEproof}
	For any SIC decoding order $(\sigma(1), \ldots, \sigma(K))$, which is a permutation of $(1,\ldots,K)$, the power allocation strategy $(\tilde{a}_{\sigma(1)}^\text{oma}, \ldots, \tilde{a}_{\sigma(K)}^\text{oma})$ such that $\mathcal{B}_{\sigma(n)}^\text{out} = \mathcal{B}_{\sigma(n)}^\text{oma},\forall n=1,\ldots,K$, is given by $\tilde{a}_{\sigma(K)}^\text{oma} = \frac{2^{R_{\sigma(K)}}-1}{2^{R_{\sigma(K)}/\tau_{\sigma(K)}}-1}$ and 
	\begin{align}
		\tilde{a}_{\sigma(n)}^\text{oma} =& \frac{2^{R_{\sigma(n)}}-1}{2^{R_{\sigma(n)}/\tau_{\sigma(n)}}-1} \\
		&+ \frac{2^{R_{\sigma(n)}}-1}{2^{R_{\sigma(n)}}}\sum_{l=n+1}^K\frac{2^{R_{\sigma(l)}}-1}{2^{R_{\sigma(l)}/\tau_{\sigma(l)}}-1}\prod_{k=n}^{l-1}2^{R_{\sigma(k)}}, \nonumber
	\end{align}
	$ n=1,\ldots,K-1$, and the sum\footnote{Note that this sum is only guaranteed to equal 1 for the decoding order $(1,...,K)$.} of this power allocation strategy is
	\begin{equation}
		\sum_{n=1}^K \tilde{a}_{\sigma(n)}^\text{oma} = \frac{2^{R_{\sigma(1)}}-1}{2^{R_{\sigma(1)}/\tau_{\sigma(1)}}-1} + \sum_{l=2}^K \frac{2^{R_{\sigma(l)}}-1}{2^{R_{\sigma(l)}/\tau_{\sigma(l)}}-1}\prod_{k=1}^{l-1}2^{R_{\sigma(k)}}. 
	\end{equation}
	Suppose a SIC decoding order has consecutive SIC decoding stages $m$ and $m+1$ such that $\sigma(m)>\sigma(m+1)$, i.e. the index the user whose signal is decoded at SIC stage $m$ is greater than the index of the user whose signal is decoded at SIC stage $m+1$, which means that $\frac{R_{\sigma(m)}}{\tau_{\sigma(m)}}>\frac{R_{\sigma(m+1)}}{\tau_{\sigma(m+1)}}$. If the SIC stages $m$ and $m+1$ are reversed, such that the signal of user $m+1$ is now decoded before the signal of user $m$, while all other stages remain in the same order, then call $(\tilde{a}_{\sigma(1)}^{'\text{oma}},\ldots,\tilde{a}_{\sigma(K)}^{'\text{oma}})$ the new power allocation strategy which has NOMA outage performance equal to OMA. It can easily be shown that $\tilde{a}_{\sigma(n)}^{'\text{oma}} = \tilde{a}_{\sigma(n)}^{\text{oma}}, \forall n\neq m, m+1$, and that $\tilde{a}_{\sigma(m)}^{'\text{oma}}$ and $\tilde{a}_{\sigma(m+1)}^{'\text{oma}}$ are given by 
	\begin{align*}
		&\tilde{a}_{\sigma(m+1)}^{'\text{oma}} = \\
		&\frac{2^{R_{\sigma(m+1)}}-1}{2^{R_{\sigma(m+1)}/\tau_{\sigma(m+1)}}-1} + (2^{R_{\sigma(m+1)}}-1)\times\bigg[\frac{2^{R_{\sigma(m)}}-1}{2^{R_{\sigma(m)}/\tau_{\sigma(m)}}-1} \\
		&+ \sum_{l=m+2}^K\frac{2^{R_{\sigma(l)}}-1}{2^{R_{\sigma(l)}/\tau_{\sigma(l)}}-1}\prod_{\substack{k=m\\k\neq m+1}}^{l-1}2^{R_{\sigma(k)}}\bigg]\hspace{-1mm},\\
		&\tilde{a}_{\sigma(m)}^{'\text{oma}} = \\ 
		&\frac{2^{R_{\sigma(m)}}-1}{2^{R_{\sigma(m)}/\tau_{\sigma(m)}}-1} + (2^{R_{\sigma(m)}}-1)\bigg[\frac{2^{R_{\sigma(m+2)}}-1}{2^{R_{\sigma(m+2)}/\tau_{\sigma(m+2)}}-1} \\
		&+ \sum_{l=m+3}^K\frac{2^{R_{\sigma(l)}}-1}{2^{R_{\sigma(l)}/\tau_{\sigma(l)}}-1}\prod_{k=m+2}^{l-1}2^{R_{\sigma(k)}}\bigg].
	\end{align*}
	Taking the difference of the sums of the two power allocation strategies gives $\sum_{n=1}^K(\tilde{a}_{\sigma(n)}^{\text{oma}} - \tilde{a}_{\sigma(n)}^{'\text{oma}})$
	\begin{align}
		=&\bigg[ \frac{(2^{R_{\sigma(m)}}-1)(2^{R_{\sigma(m+1)}}-1)}{2^{R_{\sigma(m+1)}/\tau_{\sigma(m+1)}}-1} \\
		&- \frac{(2^{R_{\sigma(m)}}-1)(2^{R_{\sigma(m+1)}}-1)}{2^{R_{\sigma(m)}/\tau_{\sigma(m)}}-1} \bigg] \prod_{k=1}^{m-1} 2^{R_{\sigma(k)}} > 0, \nonumber
	\end{align}
	which is true because $\frac{R_{\sigma(m)}}{\tau_{\sigma(m)}} > \frac{R_{\sigma(m+1)}}{\tau_{\sigma(m+1)}}$, and thus power allocation strategy $\{\tilde{a}_{\sigma(n)}^{'\text{oma}}\}_{n=1}^K$ is more energy efficient than $\{\tilde{a}_{\sigma(n)}^{\text{oma}}\}_{n=1}^K$. Successively repeat this process of reversing the positions of all consecutive SIC decoding stages $m$ and $m+1$ which have $\sigma(m)>\sigma(m+1)$, while keeping the SIC decoding order of every other stage constant, in order to successively obtain a more energy efficient power allocation strategy. This process is repeated until the SIC decoding order obtained is given by $(1,\ldots,K)$. 
\end{IEEEproof}

\subsection{Proof of Theorem \ref{thm:well-behaved-NOMA}}\label{proof:well-behaved-NOMA}
\begin{IEEEproof}
		(a) Since $a_{n}^\text{oma}=\frac{2^{R_{n}}-1}{2^{R_{n}/\tau_{n}}-1} + (2^{R_{n}}-1)(A_{n}^\text{oma}+c_n)$, and $\epsilon_{n-1}$ and $\epsilon_n=0$, then proposition \ref{prop:power_coeff_relate} is used to show that the following is always true for $n=2,\ldots,K$:
			\begin{align}
				&a_{n-1}^\text{oma} > a_n^\text{oma}2^{R_n}\frac{2^{R_{n-1}}-1}{2^{R_n}-1}\\
				&\Longleftrightarrow \nonumber\\ 
				&\frac{2^{R_{n-1}}-1}{2^{R_{n-1}/\tau_{n-1}}-1} + (2^{R_{n-1}}-1)(A_{n-1}^\text{oma}+2^{R_n}c_n)\\
				&> \left(\frac{2^{R_{n}}-1}{2^{R_{n}/\tau_n}-1} + (2^{R_{n}}-1)(A_{n}^\text{oma}+c_n)\right)2^{R_n}\frac{2^{R_{n-1}}-1}{2^{R_n}-1}\nonumber\\
				&\Longleftrightarrow \nonumber\\
				&\frac{1}{2^{R_{n-1}/\tau_{n-1}}-1} + (\tilde{a}_n^\text{oma} + A_{n}^\text{oma}) > \frac{2^{R_n}}{2^{R_{n}/\tau_{n}}-1} + 2^{R_n}A_{n}^\text{oma}\\
				&\Longleftrightarrow\frac{1}{2^{R_{n-1}/\tau_{n-1}}-1}  > \frac{1}{2^{R_{n}/\tau_{n}}-1},
			\end{align}
			which is true because $\frac{R_{n-1}}{\tau_{n-1}}<\frac{R_n}{\tau_n}$.
		
		(b) For this case, the minimum allowable power allocation coefficient for users $m=1,\ldots,n-1$ is $a_m^\text{oma}$. If $\epsilon_n>0$, user $n$ power allocation coefficient $a_n = a_n^\text{oma}+\epsilon_n$ leads to $\mathcal{B}_n\subset\mathcal{B}_n^\text{oma}$. The power allocation coefficient for user $n-1$ is then given by 
			\begin{align}
				&a_{n-1} = \frac{2^{R_{n-1}}-1}{2^{R_{n-1}/\tau_{n-1}}-1} + (2^{R_{n-1}}-1)( A_{n-1}^\text{oma} + \epsilon_n) + \epsilon_{n-1}\nonumber\\
				\label{eq:add_power_compensate1}&= \tilde{a}_{n-1}^\text{oma}+(2^{R_{n-1}}-1)\epsilon_n + \epsilon_{n-1}.
			\end{align}
			Then, the power allocation coefficient for users $m=1,\ldots,n-2$ is found recursively starting from $n-2$ to be
			\begin{equation}
				\label{eq:add_power_compensate2}a_m = \tilde{a}_{m}^\text{oma}+\epsilon_m+ (2^{R_{m}}-1)\left(\epsilon_{m+1}+\hspace{-1mm}\sum_{l=m+2}^{n}\hspace{-1mm}\epsilon_l\hspace{-1mm}\prod_{k=m+1}^{l-1}2^{R_k}\right).
			\end{equation}
			By noting that the sum of the power allocation coefficients\footnote{Note that $c_n\geq 0$, with potential equality if no additional power allocation is available} is less than or equal to $1$, 
			\begin{align}
				&1\geq \sum_{m=1}^K a_m=\sum_{m=1}^{K}\tilde{a}_m^\text{oma} + d_n+\epsilon_n\prod_{l=1}^{n-1}2^{R_l}\\
				&= A_\text{tot}^\text{oma} + d_n+\epsilon_n\prod_{l=1}^{n-1}2^{R_l}\nonumber\\
				\label{eq:epsilon_bound1}&\Rightarrow \epsilon_n\leq (1-A_\text{tot}^\text{oma})\prod_{l=1}^{n-1}2^{-R_l}-\sum_{m=1}^{n-1}\epsilon_m\prod_{l=m}^{n-1}2^{-R_l}
			\end{align}
		It must also be true that 
			\begin{align}
				&a_{n-1} \geq a_n 2^{R_n}\frac{2^{R_{n-1}}-1}{2^{R_n}-1}\Longleftrightarrow \nonumber\\
				&\frac{2^{R_{n-1}}-1}{2^{R_{n-1}/\tau_{n-1}}-1} +\epsilon_{n-1}+ (2^{R_{n-1}}-1)(A_{n-1}^\text{oma}+c_n2^{R_n}+\epsilon_n) \nonumber\\
				&\geq \bigg[\frac{2^{R_{n}}-1}{2^{R_{n}/\tau_{n}}-1} + \epsilon_n + (2^{R_{n}}-1)(A_{n}^\text{oma}+c_n)\bigg] 2^{R_n}\tfrac{2^{R_{n-1}}-1}{2^{R_n}-1}\nonumber\\
				&\Longleftrightarrow\label{eq:epsilon_bound2}\epsilon_n \leq \epsilon_{n-1}\frac{2^{R_{n}}-1}{2^{R_{n-1}}-1}+\frac{2^{R_{n}}-1}{2^{R_{n-1}/\tau_{n-1}}-1} - \frac{2^{R_{n}}-1}{2^{R_{n}/\tau_{n}}-1}.
			\end{align}
			The inequalities (\ref{eq:epsilon_bound1}) and (\ref{eq:epsilon_bound2}) hence yield the result in inequality (\ref{eq:epsilon_bound}).
\end{IEEEproof}



\section{Three-user Example Demonstrating Main Points \label{page:example} }

{\bf The following demonstrates the results in the manuscript using a three-user downlink NOMA system}. \\

Suppose that three users have been determined to be suitable by the base-station to receive downlink transmissions of signals with specific target rates, which will occur within a transmission time period of $T$ seconds. The suitability determined by the base-station is based on received CSI (e.g. RI, PMI, CQI, etc.) feedback from the users, as well as its own channel estimates based on prior uplink transmissions from these same users in the same bandwidth. Let the users be named user-$1$, user-$2$, and user-$3$. The target rate for user-$1$ is $R_1$, user-$2$ is $R_2$, and user-$3$ is $R_3$, all in terms of bits per second per Hz (bps/Hz). \\

Suppose that the base-station determines that the downlink transmissions to these users can be scheduled according to an orthogonal multiple access approach, TDMA, where the base-station allocates fractions of the transmission time $\tau_1$ to user-$1$, $\tau_2$ to user-$2$, and $\tau_3$ to user-$3$ respectively, and the signals are transmitted with full transmit SNR $\xi$ for their respective time duration. The selection of $\tau_1, \tau_2, \tau_3$ can be any set positive values such that $\tau_1 + \tau_2 + \tau_3 = 1$, even the optimum values for this system. Assume the relationship $\frac{R_1}{\tau_1}<\frac{R_2}{\tau_2}<\frac{R_3}{\tau_3}$ determined the user indices.\\

The base-station can instead schedule a NOMA downlink transmission for the signals of these same users instead, where the power allocation strategy is $(a_1, a_2, a_3)$. The base-station selects an appropriate power allocation strategy $(a_1, a_2, a_3)$ for the SIC decoding order $(1,2,3)$ such that:
\begin{itemize}
	\item the total inferference coefficients of when decoding signals $1$ and $2$ must not exceed the thresholds $A_1 = 2^{-R_1}$ and $A_2 = 2^{-(R_1+R_2)}$, respectively;
	\item the power allocation strategy is well-behaved, which means that $\mathcal{B}_n^\text{out} = \mathcal{B}_n, n=1,2,3$, and a clear example is given that a power allocation strategy that is not well-behaved is wasting transmit power without the additional gain in outage probability performance;
	\item the outage probability of this NOMA system will outperform the OMA system given above, and do so in the most energy efficient manner.
\end{itemize}
The following will demonstrate the above points regarding the NOMA power allocation strategy by first deriving the power allocation coefficients. \\

{\bf Total Interference of signals:} If the power allocation strategy is such that when decoding the first signal, the total interference coefficient received $A_1=a_2+a_3>2^{-R_1}$, then the outage event of decoding this signal at the receiver of user-$n=1,2,3$, is such that $\log_2(1+\frac{a_1\xi|G_n|^2}{1+A_1\xi|G_n|^2})<R_1$. So solving for the channel SNR gain $|G_n|^2$ gives
	\begin{align*}
		&a_1|G_n|^2<(2^{R_1}-1)(1+A_1\xi|G_n|^2)\\
		\Longrightarrow& \xi|G_n|^2 [a_1 - A_1(2^{R_1}-1)] < 2^{R_1}-1\\
		\Longrightarrow& \xi|G_n|^2 [1- A_12^{R_1}] < 2^{R_1}-1\\
	\end{align*}
Since $A_1>2^{-R_1}\Rightarrow A_12^{R_1} > 1$, so $1-A_12^{R_1} < 1$. Therefore, the following is true
	\begin{align*}
		\Longrightarrow \frac{2^{R_1}-1}{\xi[1- A_12^{R_1}]} < 0 < |G_n|^2 .
	\end{align*}
which means that there is an outage event $\forall |G_n|^2 > 0$ (indeed, all channel SNR gains are positive real numbers). Therefore, the interference coefficient $A_1 = a_2+a_3<R_1$ must hold true in order to not create a {\em certain outage}. In the same manner, if interference coefficient $A_2 = a_3 > 2^{-(R_1+R_2)}$, since the outage event at the receiver of user-$n=2,3$ is given by $\log_2(1+\frac{a_2\xi|G_n|^2}{1+A_2\xi|G_n|^2})<R_2$, then 
	\begin{align*}
		\Longrightarrow& \xi|G_n|^2[a_2-A_2(2^{R_2}-1)] < 2^{R_2}-1\\
		\Longrightarrow& \xi|G_n|^2[A_1-A_2 2^{R_2}] < 2^{R_2}-1.
	\end{align*}
If $A_1 = a_2+a_3 <2^{-R_1}$ then $A_1 - A_2 2^{R_2} < 0$
	\begin{align*}
		\Longrightarrow&  \frac{2^{R_2}-1}{\xi[A_1-A_2 2^{R_2}]}<0<|G_n|^2,
	\end{align*}
Otherwise if $A_1>2^{-R_1}$ then stage 1 of the SIC decoding procedure will fail at all users.\\

{\bf Well-behaved Property:} The power allocation strategy should be such that the coefficients have the property $a_1\geq a_2\frac{2^{R_2}(2^{R_1}-1)}{2^{R_2}-1}$ and $a_2\geq a_3\frac{2^{R_3}(2^{R_2}-1)}{2^{R_3}-1}$ so that the $\mathcal{B}_2^\text{out} = \mathcal{B}_2\cup\mathcal{B}_{2\rightarrow 1} = \mathcal{B}_2$ and $\mathcal{B}_3^\text{out} = \mathcal{B}_3\cup\mathcal{B}_{3\rightarrow 1}\cup\mathcal{B}_{3\rightarrow 2} = \mathcal{B}_3$. This can be demonstrated in the proof of proposition 1 to ensure that the interference coefficients do not exceed the upper-bounds described in theorem 1. \\

Also, this is important because of the following example. Let $a_3 = \tilde{a}_3^\text{oma}$, $a_2 = \tilde{a}_2^\text{oma} + \epsilon_2$ and $a_1 = \tilde{a}_1^\text{oma} + (2^{R_1}-1)\epsilon_2$, such that $a_1+a_2+a_3=1$ and $\epsilon_2>\frac{2^{R_2}-1}{2^{R_1/\tau_1}-1}-\frac{2^{R_2}-1}{2^{R_2/\tau_2}-1}$. This leads to $a_1<a_2\frac{2^{R_2}(2^{R_1}-1)}{2^{R_2}-1}$ and $a_2\geq a_3\frac{2^{R_3}(2^{R_2}-1)}{2^{R_3}-1}$, so then $\mathcal{B}_1^\text{out} = \mathcal{B}_1^\text{oma}$ and $\mathcal{B}_3^\text{out} = \mathcal{B}_3^\text{oma}$. However, $\mathcal{B}_2^\text{out} = \mathcal{B}_{2\rightarrow 1}$, because as is shown in the proof of theorem 3, and since $\mathcal{B}_2 = \left\{ |G_2|^2 < \frac{2^{R_2}-1}{\xi[a_2 - a_3(2^{R_2}-1)]} \right\}, \mathcal{B}_{2\rightarrow 1} = \left\{ |G_2|^2 < \frac{2^{R_1}-1}{\xi[a_1 - (a_2+a_3)(2^{R_1}-1)]} \right\}$ and 
	\begin{align*}
		&\frac{2^{R_2}-1}{\xi[a_2 - a_3(2^{R_2}-1)]}< \frac{2^{R_1}-1}{\xi[a_1 - (a_2+a_3)(2^{R_1}-1)]} \\
		&\Longrightarrow\mathcal{B}_{2}\subset\mathcal{B}_{2\rightarrow 1}\\
	\end{align*}
$\Longrightarrow \mathcal{B}_2^\text{out}=\mathcal{B}_2\cup\mathcal{B}_{2\rightarrow 1}=\mathcal{B}_{2\rightarrow 1}$. \\

In plain terms, this means that the power allocation for user-$2$ can be reduced to the level such that $\epsilon_2=\frac{2^{R_2}-1}{2^{R_1/\tau_1}-1}-\frac{2^{R_2}-1}{2^{R_2/\tau_2}-1}$, which leads to $\mathcal{B}_{2}=\mathcal{B}_{2\rightarrow 1}$, meaning that the power allocation for user-$2$ can be reduced, while maintaining the same power allocation for users-$1$ and user-$3$, and still achieve the exact outage probability performance for all 3 users. In other words, a power allocation strategy that is not well-behaved is wasting power on one of the signals when compared to another existing power allocation strategy that achieves the same outage probability performance for all users.\\

{\bf Power Allocation so that NOMA has Equal Outage Performance to OMA for all Users:} For the three users, the power allocation strategy that will outage probability performance equal to OMA (and to have sum less than 1 in the proof of theorem 2, and to be well-behaved and not violate the interference property in theorem 3) is given by 
	\begin{align*}
		&\tilde{a}_1 = \frac{2^{R_1}-1}{2^{R_1/\tau_1}-1} + (2^{R_1}-1)\hspace{-1mm}\left(\hspace{-0.5mm}\frac{2^{R_2}-1}{2^{R_2/\tau_2}-1} + 2^{R_2}\frac{2^{R_3}-1}{2^{R_3/\tau_3}-1}\hspace{-0.5mm}\right)\\
		&\tilde{a}_2 = \frac{2^{R_2}-1}{2^{R_2/\tau_2}-1} + (2^{R_2}-1)\frac{2^{R_3}-1}{2^{R_3/\tau_3}-1}\\
		&\tilde{a}_3 = \frac{2^{R_3}-1}{2^{R_3/\tau_3}-1}.
	\end{align*}
The summation of this power allocation strategy can be shown to be less than 1 as follows. The indices are assigned based on the relationship of $\frac{R_1}{\tau_1} < \frac{R_2}{\tau_2} < \frac{R_3}{\tau_3}$. Define $\tau_1=\frac{b_1}{K}, \tau_2=\frac{b_2}{K}, \tau_3=\frac{b_3}{K}$, then $\frac{R_1}{\tau_1} = \frac{KR_1}{b_1}, \frac{R_2}{\tau_2} = \frac{KR_2}{b_2}, \frac{R_3}{\tau_3} = \frac{KR_3}{b_3}$. Defining $r_1 = \frac{R_1}{b_1}, r_2 = \frac{R_2}{b_2}, r_3 = \frac{R_3}{b_3}$, clearly $r_1<r_2<r_3$. Therefore, since $f(t) = \frac{2^{bt}-1}{2^{Kt}-1}, b<K,$ is monotonically decreasing in $t$, then
	\begin{align*}
		&a_1+a_2+a_3 \\
		&= \frac{2^{R_1}-1}{2^{R_1/\tau_1}-1} + 2^{R_1}\frac{2^{R_2}-1}{2^{R_2/\tau_2}-1} + 2^{R_1+R_2}\frac{2^{R_3}-1}{2^{R_3/\tau_3}-1}\\
		&= \frac{2^{b_1r_1}-1}{2^{Kr_1}-1} + 2^{b_1r_1}\frac{2^{b_2r_2}-1}{2^{Kr_2}-1} + 2^{b_1r_1+b_2r_2}\frac{2^{b_3r_3}-1}{2^{Kr_3}-1}\\
		&< \frac{2^{b_1r_1}-1}{2^{Kr_1}-1} + 2^{b_1r_1}\frac{2^{b_2r_2}-1}{2^{Kr_2}-1} + 2^{b_1r_1+b_2r_2}\frac{2^{b_3r_2}-1}{2^{Kr_2}-1}\\
		&= \frac{2^{b_1r_1}-1}{2^{Kr_1}-1} + 2^{b_1r_1}\frac{2^{(b_2+b_3)r_2}-1}{2^{Kr_2}-1} \\
		&< \frac{2^{b_1r_1}-1}{2^{Kr_1}-1} + 2^{b_1r_1}\frac{2^{(b_2+b_3)r_1}-1}{2^{Kr_1}-1} \\
		&= \frac{2^{(b_1+b_2+b_3)r_1}-1}{2^{Kr_1}-1}\\
		&=1.
	\end{align*}

{\bf Energy Efficiency of Proposed SIC Decoding Order:} Corollary 1 states that if a different SIC decoding order $(\sigma(1), \sigma(2), \sigma(3))$ is used, where $(\sigma(1), \sigma(2), \sigma(3))$ is any permutation of the sequence $(1,2,3)$, then a more energy efficient SIC decoding order can be found if $\sigma(1)>\sigma(2)$ or $\sigma(2)<\sigma(3)$, by switching the SIC decoding order. Then, the power allocation strategy that enables NOMA to achieve equal outage performance to OMA for each user is given by 
	\begin{align*}
		&\tilde{a}_{\sigma(1)} = \frac{2^{R_{\sigma(1)}}-1}{2^{R_{\sigma(1)}/\tau_{\sigma(1)}}-1} \\
		&+ (2^{R_{\sigma(1)}}-1)\left(\frac{2^{R_{\sigma(2)}}-1}{2^{R_{\sigma(2)}/\tau_{\sigma(2)}}-1} + 2^{R_{\sigma(2)}}\frac{2^{R_{\sigma(3)}}-1}{2^{R_{\sigma(3)}/\tau_{\sigma(3)}}-1}\right)\\
		&\tilde{a}_{\sigma(2)} = \frac{2^{R_{\sigma(2)}}-1}{2^{R_{\sigma(2)}/\tau_{\sigma(2)}}-1} + (2^{R_{\sigma(2)}}-1)\frac{2^{R_{\sigma(3)}}-1}{2^{R_{\sigma(3)}/\tau_{\sigma(3)}}-1}\\
		&\tilde{a}_{\sigma(3)} = \frac{2^{R_{\sigma(3)}}-1}{2^{R_{\sigma(3)}/\tau_{\sigma(3)}}-1}.
	\end{align*}
	The following permutations $(\sigma(1), \sigma(2), \sigma(3))$ of $(1,2,3)$ exist and can be the SIC decoding order for the system: $(2,1,3), (2,3,1), (1,3,2),(3,1,2),(3,2,1),$ and the identity permutation $(1,2,3)$. By the proof of corollary 1, the SIC decoding orders can be listed in order of least energy efficient to most energy efficient: $\{(3,2,1),(3,1,2),(1,3,2),(1,2,3)\}$ or $\{(3,2,1),(2,3,1),(2,1,3),(1,2,3)\}$ For example, the proof of corollary 1 can be used to show that the SIC decoding order $(3,1,2)$ is more energy efficient (i.e. uses less transmit power) than $(3,2,1)$ in order to enable NOMA to achieve equal outage probability performance to OMA. Call $S_{(3,2,1)}$ and $S_{(3,1,2)}$ the sums of the power allocation strategies given above for the SIC decoding orders $(3,2,1)$ and $(3,1,2)$, respectively. Then according to the proof of corollary 1 it is shown that $S_{(3,2,1)}-S_{(3,1,2)}>0$ as follows
	\begin{align*}
		&S_{(3,2,1)} = \frac{2^{R_3}-1}{2^{R_3/\tau_3}-1} + 2^{R_3}\frac{2^{R_2}-1}{2^{R_2/\tau_2}-1} + 2^{R_2+R_3}\frac{2^{R_1}-1}{2^{R_1/\tau_1}-1},\\
		&S_{(3,1,2)} = \frac{2^{R_3}-1}{2^{R_3/\tau_3}-1} + 2^{R_3}\frac{2^{R_1}-1}{2^{R_1/\tau_1}-1} + 2^{R_1+R_3}\frac{2^{R_2}-1}{2^{R_2/\tau_2}-1}\\
		&\Longrightarrow S_{(3,2,1)}-S_{(3,1,2)} = \frac{2^{R_3}-1}{2^{R_3/\tau_3}-1} + 2^{R_3}\frac{2^{R_2}-1}{2^{R_2/\tau_2}-1} \\
		& + 2^{R_2+R_3}\frac{2^{R_1}-1}{2^{R_1/\tau_1}-1} - \frac{2^{R_3}-1}{2^{R_3/\tau_3}-1} - 2^{R_3}\frac{2^{R_1}-1}{2^{R_1/\tau_1}-1} \\
		& - 2^{R_1+R_3}\frac{2^{R_2}-1}{2^{R_2/\tau_2}-1}\\
		&			= 2^{R_3}(2^{R_1}-1)(2^{R_2}-1)\left(\frac{1}{2^{R_1/\tau_1}-1}-\frac{1}{2^{R_2/\tau_2}-1}\right)\\
		&			> 0,
	\end{align*}
which is true because $\frac{R_1}{\tau_1}<\frac{R_2}{\tau_2}$. Using the same steps, it is easy to show that $S_{(3,1,2)}-S_{(1,3,2)}>0$ and $S_{(1,3,2)}-S_{(1,2,3)}>0$. Therefore, it is clear that the SIC decoding order $(1,2,3)$ is more energy efficient that any of the possible SIC decoding orders in terms of the power allocation required for NOMA to achieve the same outage probility performance as OMA. Again, it should be noted that the sum of the power allocation strategies $(\tilde{a}_{\sigma(1)}, \tilde{a}_{\sigma(2)},\tilde{a}_{\sigma(3)})$ is not guaranteed to be less than 1 for all other SIC decoding orders other than $(1,2,3)$, as was proven in theorem 2. Nonetheless, even if there are other SIC decoding orders with power allocation strategies $(\tilde{a}_{\sigma(1)}, \tilde{a}_{\sigma(2)},\tilde{a}_{\sigma(3)})$ such that the sum is less than 1, the power headroom available in order to increase the power allocation beyond $(\tilde{a}_{\sigma(1)}, \tilde{a}_{\sigma(2)},\tilde{a}_{\sigma(3)})$ is still less for all SIC decoding orders than that available for $(1,2,3)$, and thus the improvement of outage probability performance of NOMA over OMA is less than for the case of the all other SIC decoding orders.\\

{\bf Power Allocation Strategy for NOMA to outperform Outage Probability of OMA:} The last matter in this example is to demonstrate how to construct well-behaved power allocation strategy which allows NOMA to outperform the outage probability performance of OMA for all users, based on all the above results. After theorem 3, a simple approach is used based on the property of well-behaved power allocation (although not necessarily optimal). For the three-user case, this power allocation approach is given by calling $A_\text{tot}^\text{oma} = \tilde{a}_1^\text{oma}+\tilde{a}_2^\text{oma}+\tilde{a}_3^\text{oma}$, and then according to equation (25) in the paper
	\begin{align*}
		\epsilon_1 & = (1-A_\text{tot}^\text{oma})\frac{2^{R_1}-1}{2^{R_1+R_2+R_3}-1}2^{R_2+R_3},\\
		\epsilon_2 & = (1-A_\text{tot}^\text{oma})\frac{2^{R_2}-1}{2^{R_1+R_2+R_3}-1}2^{-R_1+R_3}\\
		\epsilon_3 & = (1-A_\text{tot}^\text{oma})\frac{2^{R_3}-1}{2^{R_1+R_2+R_3}-1}2^{-R_1-R_2}.
	\end{align*}
As shown in the paper and verified in the simulation results, this additional power allocation combined with the general description of the power allocation coefficients in equation (17) provide a simple framework for vastly improving the outage performance of downlink NOMA over OMA, without complex suboptimal searches or relying on the base-station having exact knowledge of all of the channel gains.

\end{document}